\documentclass[12pt]{article}
\usepackage[totalwidth=470truept,totalheight=636truept]{geometry}
\usepackage{latexsym,graphicx,amsfonts,amssymb,amsmath,color,dsfont,cancel,tikz,multirow,slashed}
\usepackage[hypertexnames=false,hidelinks]{hyperref}
\usepackage{diagbox}
\usepackage{slashbox}
\usepackage[export]{adjustbox}
\usepackage[format=hang,font=small,textfont=it]{caption}

\newcommand{\AB}{\includegraphics[width=0.5in]{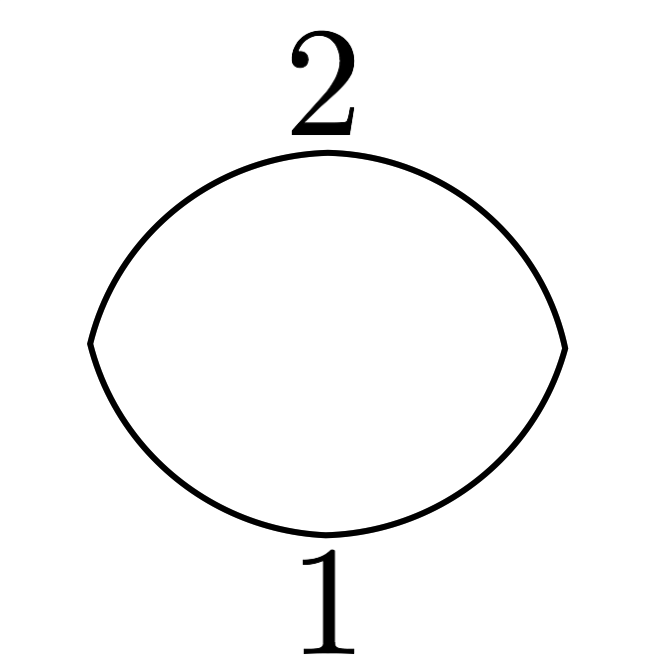}}
\newcommand{\ApBp}{\includegraphics[width=0.5in]{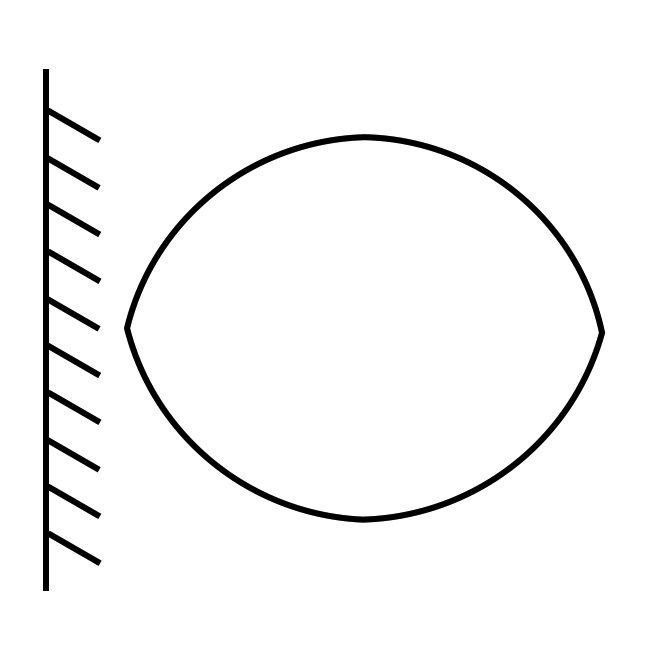}}
\newcommand{\ApB}{\includegraphics[width=0.5in]{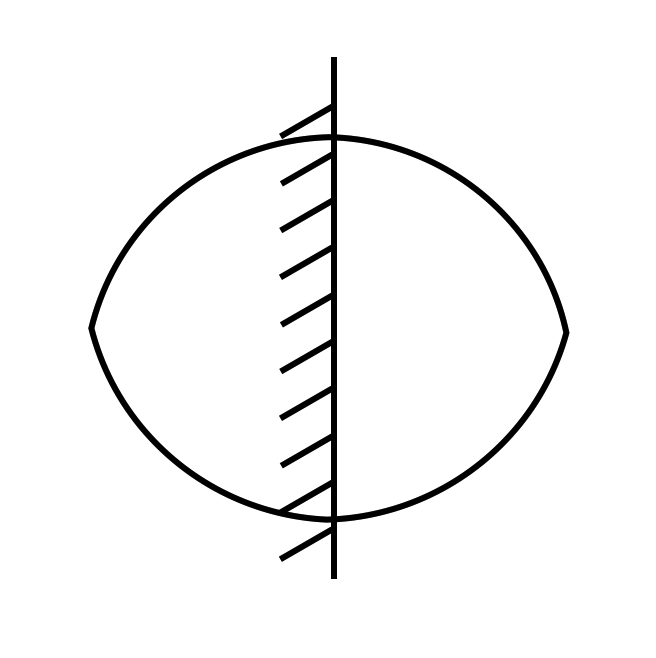}}
\newcommand{\ABp}{\includegraphics[width=0.5in]{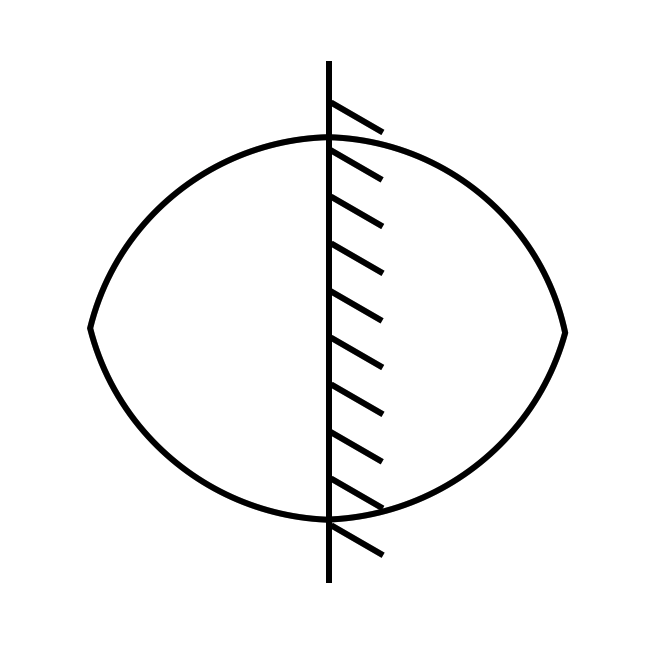}}

\newcommand{\ABC}{\includegraphics[width=0.48in]{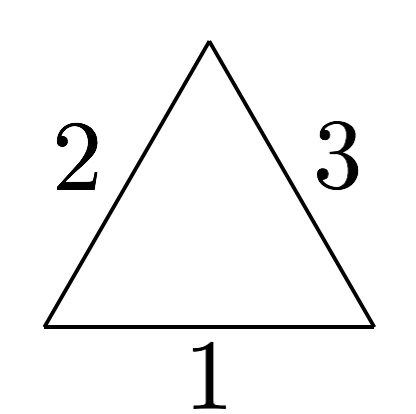}}
\newcommand{\ApBpCp}{\includegraphics[width=0.48in]{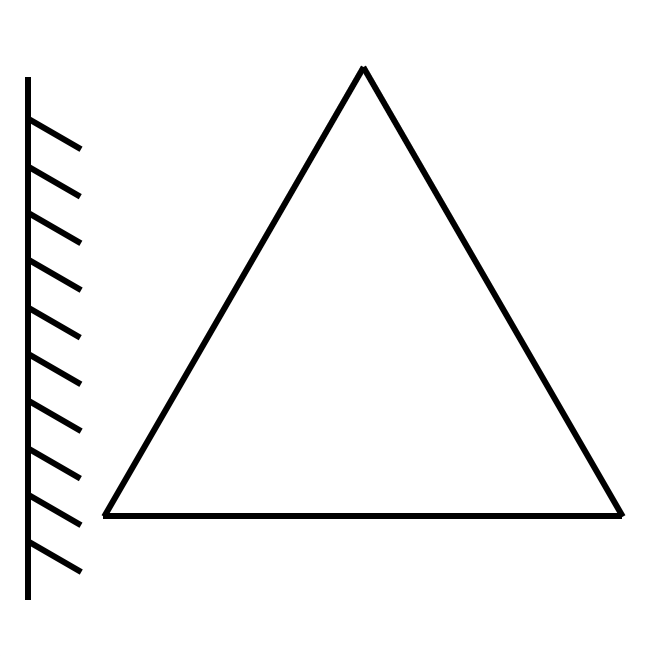}}
\newcommand{\ApBC}{\includegraphics[width=0.48in]{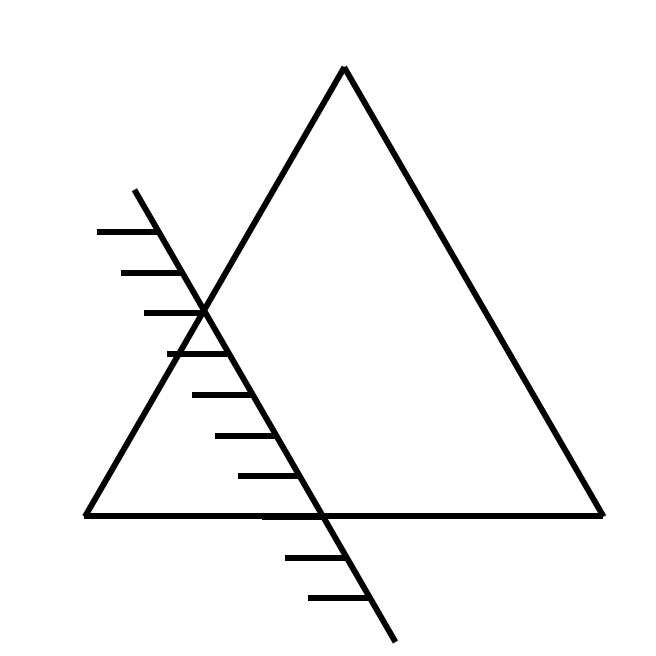}}
\newcommand{\ABpC}{\includegraphics[width=0.48in]{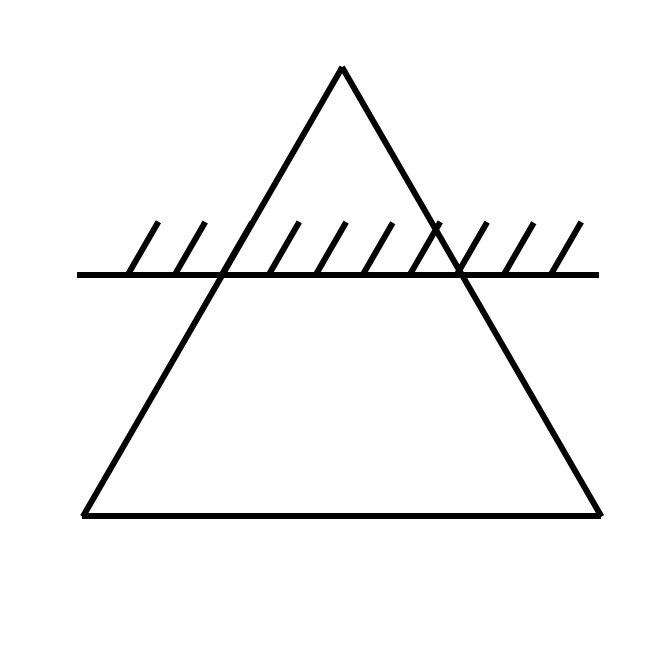}}
\newcommand{\ABCp}{\includegraphics[width=0.48in]{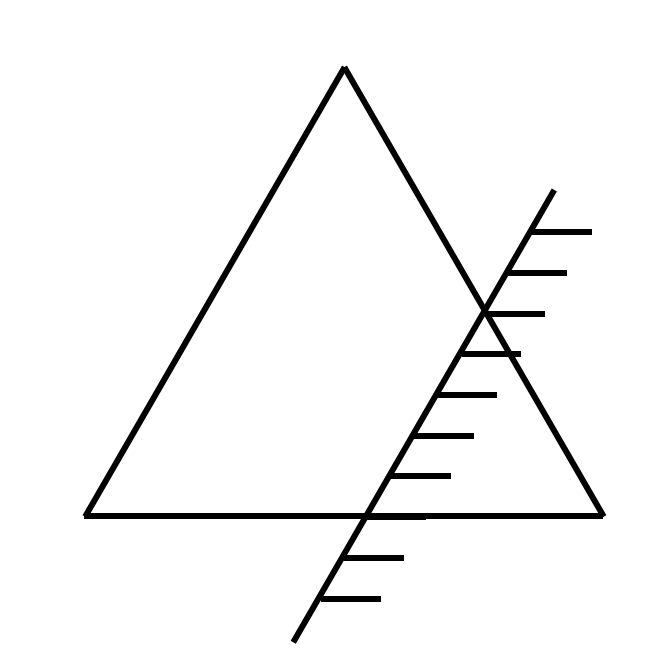}}
\newcommand{\ABpCp}{\includegraphics[width=0.48in]{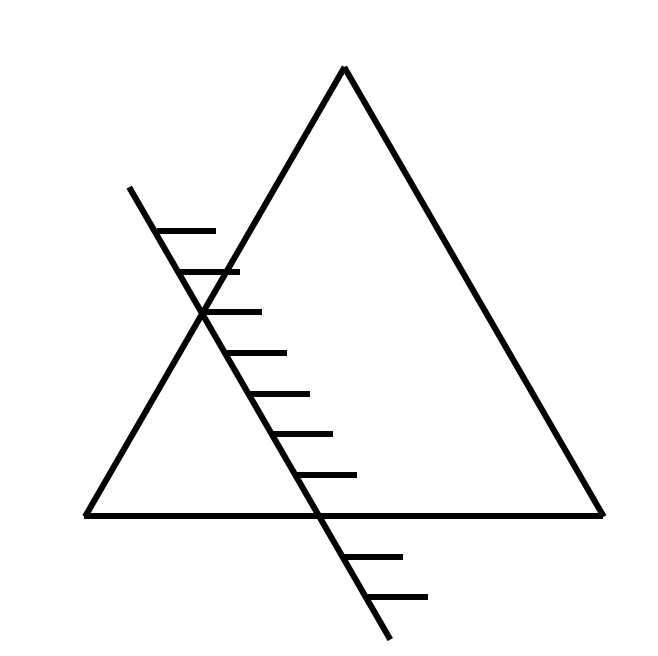}}
\newcommand{\ApBCp}{\includegraphics[width=0.48in]{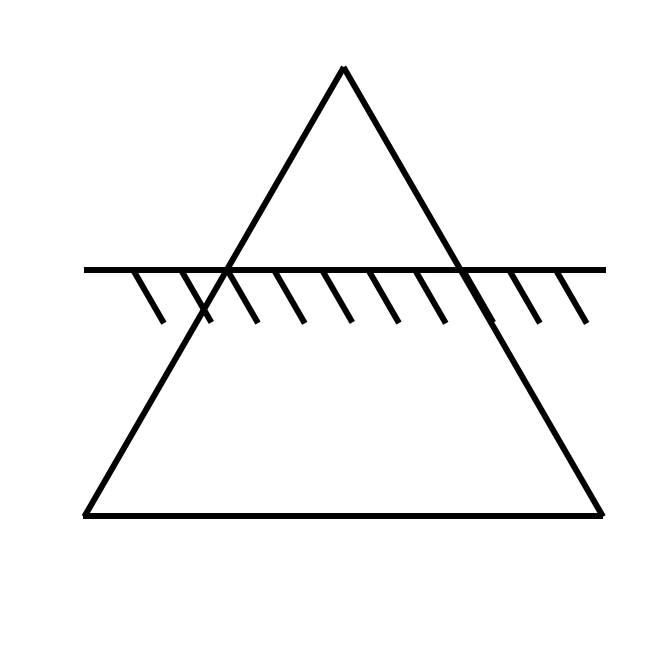}}
\newcommand{\ApBpC}{\includegraphics[width=0.48in]{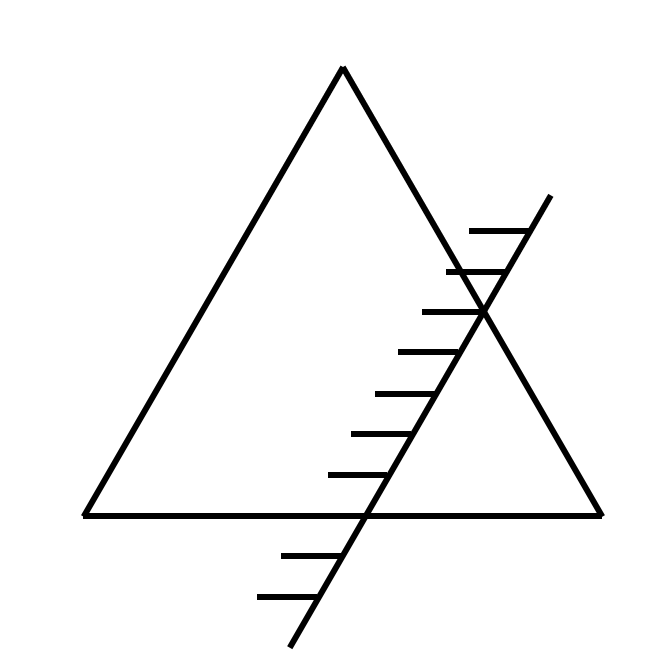}}

\DeclareMathAlphabet{\mymathbb}{U}{BOONDOX-ds}{m}{n}

\newcommand{\sect}[1]{\setcounter{equation}{0}\section{#1}}

\global\arraycolsep=1truept

\hypersetup{colorlinks=true,
    linkcolor=black,
    filecolor=black, 
    citecolor=black,
    urlcolor=blue}

\RequirePackage{booktabs}

\begin{document}

\null

\vskip1truecm

\begin{center}
{\LARGE \textbf{Higher-Derivative Quantum Gravity}}

\vskip.6truecm

{\LARGE \textbf{With Purely Virtual Particles:}}

\vskip.6truecm

{\LARGE \textbf{Renormalizability And Unitarity}}

\vskip1truecm

\textsl{{\large Marco Piva}}

\vskip .1truecm

{\textit{Faculty of Physics, University of Warsaw,\\ Pasteura 5, 02-093 Warsaw, Poland}}

\vspace{0.2cm}

mpiva@fuw.edu.pl
\vskip1truecm

\textbf{Abstract}
\end{center}
We review the formulation of quantum field theories with purely virtual particles, a new type of degrees of freedom that can mediate interactions without ever appear as external on-shell states. This property allows to solve the problem of ghosts in higher-derivative quantum gravity, leading to a renormalizable and unitary theory. The main steps for the BRST quantization of gravity are recalled and renormalizability is discussed. Then, we introduce purely virtual particles in a general quantum field theory and show the derivation of the so-called spectral identities, which are a key ingredient to prove unitarity. Finally, phenomenological consequences and predictions in inflationary cosmology are presented.
\vfill\eject

\sect{Introduction}
The formulation of a theory of quantum gravity is one of the most important task in theoretical physics. It is expected to shed light on several open problems and, most importantly, it will tell us something new about our understanding of reality, in the same way quantum mechanics made many of our certainties crumble at the beginning of the twentieth century. Many phenomenological questions, such as the understanding of the initial phases of the universe or how to deal with black holes in quantum mechanics, as well as more fundamental ones (do the notions of time, cause, past and future survive at arbitrary scales?) might find answer in the theory of quantum gravity. However, formulating such a theory is as important as it is hard. We will probably need many decades or centuries before to actually answer to all the relevant questions, because it is a very intricate subject and it connects many aspects of physics altogether. Moreover, we do not have enough experimental data to infer much about the quantum nature of gravity. Therefore, in the situation where the complexity of the topic comes together with the lack of data, it is important to stay with our foot on the ground and proceed carefully. The starting point is to look at the past, analyze what has worked so far and try to deviate as least as possible from that path. The risk is to completely loose contact with nature and go astray forever. 

Over the last decades there have been several proposals to tackle the problem of quantum gravity. Some of them rely on abandoning the framework that has been very successful so far, i.e. quantum field theory, in favor of something else. In some cases the approach is based on the assumption that gravity is the geometry of spacetime at arbitrary small scales. We have no evidence for this, therefore those theories (loop quantum gravity, causal dynamical triangulations, spinfoam to name a few) should be viewed as theories of quantum geometry, rather than theories of quantum gravity. Only the contact with nature will tell us whether that assumption is correct or not. We think that a more cautious approach is to study the gravitational field in quantum mechanical terms, as we do for all other fields, in the framework of quantum field theory.

In this review we present an approach to do so in a way that is consistent with the principles that have led to the standard model of particle physics: locality, renormalizability and unitarity. Although conservative, this proposal cannot avoid to introduce something new, albeit minimal. The new idea is the possibility to describe quanta that are purely virtual, i.e. they can be responsible for interactions between other particles, like a virtual photon is responsible for the interaction between two electrons, but cannot appear as on-shell states. These new type of particles are called \emph{purely virtual particles} or \emph{fakeons}, and they are introduced by means of a specific procedure that we call \emph{fakeon prescription}~\cite{Anselmi:2017yux,Anselmi:2017ygm}. Their presence is enough to have a local, renormalizable and unitary quantum field theory of gravity.

The approach presented here has its roots in results that were known since many decades. The starting point is that Einstein gravity treated as a quantum field theory is nonrenormalizable~\cite{tHooft:1974toh, Goroff:1985th}. The problem with nonrenormalizability is that, in order to reabsorb the divergences of the theory, we need to introduce infinitely many operators of increasing dimension, each of them with a new independent parameter. Therefore, in principle, if we want to fix all parameters with experiments we would need infinitely many measurements. However, all the known nonrenormalizable theories are effective theories, which means that their nonrenormalizability is due to the fact that some additional (heavy) fields are integrated out. Once those fields are properly included the infinitely many parameters turn out to be functions of a finite subset of them, and the theory is renormalizable. This is not what happens in Einstein gravity, at least not in the same way as in other theories, like the Fermi model~\cite{Fermi:1933jpa}. 

One way of proceed is to just accept that there might exist fundamental theories that are nonrenormalizable and start to work out their properties in order to make them as predictive as they can (see e.g.~\cite{Anselmi:2005vk, Anselmi:2002ge, Anselmi:2013wha}). Along this line is the asymptotic safety program~\cite{Weinberg:1980gg, Lauscher:2001ya}, which is a nonperturbative field-theoretic approach that accounts for a generalization of asymptotic freedom, where the coupling constant is assumed to flows to a nonvanishing value in the ultraviolet limit. Then, for the theory to be predictive, such a interacting fixed point needs to have a finite-dimensional critical surface. An asymptotically safe theory is sometimes referred as nonperturbatively renormalizable.

Another possibility is to go beyond Einstein gravity by imposing renormalizability. The natural choice is to add to the classical action the counterterms generated at one loop in Einstein gravity, assume that their couplings are large and redo the analysis. This was done in 1976 by Stelle~\cite{Stelle:1976gc}, showing that adding to the action of gravity all the independent terms quadratic in the Riemann tensor, as well as the cosmological term, makes the theory renormalizable. Moreover, some additional degrees of freedom are introduced in this way. Besides the usual graviton modes, one massive scalar and one massive spin-2 particles also propagate. However, the kinetic term of the latter has the wrong sign, causing instabilities that at quantum level can be traded for violations of unitarity. This type of particles are called \emph{ghosts}. At that point the theory, which we call \emph{Stelle gravity}, was (righteously) labeled as physically unacceptable and the research in this direction essentially stopped. Instead of insisting on the principle of renormalizability and try to solve the problem of ghosts, many physicists moved to something else, often different from quantum field theory, to explain quantum gravity. Over the decades only a limited number of people continued to study Stelle gravity and work out its properties, but they often postpone the problem of ghosts for later studies or invoke some nonperturbative effects or theories beyond quantum field theory that will eventually explain the matter~\cite{Julve:1978xn,Fradkin:1981iu, Kawasaki:1981gk, Tomboulis:1983sw, Avramidi:1985ki}. These explanations never came. Furthermore, in 1980, Starobinsky shown that the scalar degree of freedom introduced by the $R^2$ term can explain the accelerated expansion of the early universe and solve a number of phenomenological issues~\cite{Starobinsky:1980te}. This is one of the most efficient and simple model of inflation, called \emph{Starobinsky model}, which is included in Stelle gravity. Moreover, since also the cosmological term is generated by renormalization, dark energy might be explained within Stelle gravity. 

Despite these results in favor of Stelle gravity, addressing the problem of ghosts has never been the priority of any of the major research lines. On top of this, several potential approaches were already present in the literature (and actually inspired ours), although not all of them were thought for gravity. For example, the Lee-Wick models~\cite{Lee:1969fy,Lee:1970iw} have been used in the context of quantum gravity only in the recent years~\cite{Modesto:2015ozb,Modesto:2016ofr}, or the so-called shadow states~\cite{Nelson:1972vbp, Sudarshan:1972mwu,Nelson:1972zq}, introduced in 1972, which also share some properties with fakeons, were soon ignored. Finally, another approach to the ghost problem is to formulate quantum gravity as a particular type of nonlocal quantum field theory~\cite{Kuzmin:1989sp}. There is a decent amount of literature in this topic, but only in the last 10-15 years and only by small groups (see~\cite{Modesto:2017sdr} and references therein). 
Finally, we mention a more recent approach, introduced in~\cite{Donoghue:2019fcb}, which uses the fact that the ghost in Stelle theory is unstable to prove unitarity in situations where it has already decayed, using a generalization of Veltman’s argument of~\cite{Veltman:1963th}. The result is an effective theory, unitary at energy scales where the width of the ghost can be considered large. However, this is far from truly removing the ghost from the spectrum, since it is always possible to find energy ranges where it is long lived and cannot be ignored.

In a nutshell, many interesting ideas to tackle the problem of ghosts were already on the table since a few decades, but each of them has been either abandoned, ignored or poorly studied.

To a relatively young physicist, like the author of this review, the feeling that the historical situation just described gives is that someone did not do their job for fifty years. 

Why abandon a successful framework such as quantum field theory? Why (almost) completely ignore the existence of a renormalizable, although not unitary, theory of gravity and move to other approaches? Why pretend not to see that the best inflationary model points in the direction of that renormalizable theory? Probably the answers to these questions lie more on the social aspects of physics rather than the scientific ones. We do not try to give our personal answers in this review, so we do not hurt the feelings of anyone.

Finally, we want to clarify that it might well be that something beyond quantum field theory is necessary to achieve a more fundamental understanding of quantum gravity. However, our opinion is that no stone should be left unturned in quantum field theory before abandoning it and the topic addressed in this review tells us that this was not the case. Furthermore, as stressed above, in the lack of guidance given by nature, we cannot afford to jump into the unknown without any grasp on experimental data. We need to stick to what was successful so far and see where it leads. 

The paper is organized as follows. In~\autoref{sec:Stelle} we review Stelle gravity, discuss its renormalizability and introduce the ghost problem. In~\autoref{sec:unit} we recall the basics of unitarity in general quantum field theories, while in~\autoref{sec:spectid} we introduce the spectral identities that are necessary to prove unitarity in theories with purely virtual particles. In \autoref{sec:QGcosm} we explain how to apply the idea of purely virtual particles to Stelle gravity and show its predictions in the context of inflationary cosmology. Finally,~\autoref{sec:concl} contains our conclusions.

\emph{Notation and conventions:} We use the signature $(+,-,-,-)$ for the metric tensor. The Riemann and Ricci tensors are defined as $R^{\mu}_{ \ \nu\rho\sigma}=\partial_{\rho}\Gamma^{\mu}_{\nu\sigma}-\partial_{\sigma}\Gamma^{\mu}_{\nu\rho}+\Gamma^{\mu}_{\alpha\rho}\Gamma^{\alpha}_{\nu\sigma}-\Gamma^{\mu}_{\alpha\sigma}\Gamma^{\alpha}_{\nu\rho}$ and $R_{\mu\nu}=R^{\rho}_{ \ \mu\rho\nu}$, respectively. We write the four-dimensional integrals over spacetime points of a function $F$ of a field $\phi(x)$ as $\int\sqrt{-g}F(\phi)\equiv\int\text{d}^4x\sqrt{-g(x)} F\left(\phi(x)\right)$. We always assume that the integral of the Gauss-Bonnet term vanish, i.e. $\int\sqrt{-g}\left(R_{\mu\nu\rho\sigma}R^{\mu\nu\rho\sigma}-4R_{\mu\nu}R^{\mu\nu}+R^2\right)=0.$

\sect{Stelle gravity}\label{sec:Stelle}
In this section we introduce Stelle gravity by recalling the general method for the quantization of gravity in quantum field theory. Moreover, we study the propagator of the theory in order to show the presence of the spin-2 ghost. Finally, we give some details about renormalizability. 

The Stelle action is
\begin{equation}
S_{\text{HD}}=-\frac{1}{2\kappa ^{2}}\int \sqrt{-g}\left[
2\Lambda _{C}+\zeta R+\frac{\alpha}{2}C^2-\frac{\xi}{6}R^{2}\right] ,  \label{lhd}
\end{equation}%
where $C^2=C_{\mu\nu\rho\sigma}C^{\mu\nu\rho\sigma}$ is the squared of the Weyl tensor and $\Lambda_C$, $\zeta$, $\alpha$, $\xi$ and $\kappa$ are real parameters with dimension
\begin{equation}
    [\Lambda_C]=4, \qquad [\zeta]=2, \qquad [\alpha]=0, \qquad [\xi]=0, \qquad [\kappa]=0.
\end{equation}
Note that the parameter $\kappa$ is redundant, since $\zeta$ already accounts for the Planck mass, but it is useful to keep track of loop orders in the perturbative expansion. We can see that no parameters of negative dimension appear, even after the correct normalization of the kinetic terms. In fact, if we expand the metric around Minkowski spacetime as
\begin{equation}\label{eq:metricsplit}
    g_{\mu\nu}=\eta_{\mu\nu}+2\kappa h_{\mu\nu},
\end{equation}
where $h_{\mu\nu}$ is the graviton field, the quadratic part of the action contains terms with four derivatives that are multiplied by dimensionless parameters. Those terms are dominant in the ultraviolet and after properly normalizing them, the total action still contains parameters with non-negative dimension. In the case of Einstein gravity the four-derivative terms are absent and the kinetic term is multiplied by $\zeta$, which has positive dimension. Once that term is normalized, all the interactions contain powers of $1/\zeta$. Since the counterterms are polynomials in the parameters of the theory (with properly normalized kinetic term), then the counterterms in Einstein gravity are polynomial in $1/\zeta$, which is a parameter with negative dimension. Therefore, it is possible to build counterterms with arbitrary mass dimension, since it can always be compensated by powers of $1/\zeta$. This is why general relativity is said to be nonrenormalizable by power counting. However, to prove the renormalizability of Stelle gravity, we need to show that the counterterms satisfies certain properties, as we explain in the next subsection.

 Before to proceed with renormalizability we derive the graviton propagator and study its poles and residues. In this way we establish some notation and review the standard steps for the quantization of gravity. First, we recall that the Becchi-Rouet-Stora-Tyutin (BRST) transformations associated to diffeomorphisms read
\begin{eqnarray}\label{brststelle}
sg_{\mu \nu}&=&-\partial_{\mu}C^{\alpha}g_{\alpha\nu}-\partial_{\nu}C^{\alpha}g_{\mu\alpha}-C^{\alpha}\partial_{\alpha}g_{\mu\nu}\nonumber\\
sC^{\rho}&=&-C^{\sigma}\partial_{\sigma}C^{\rho}\nonumber\\
s\bar{C}^{\sigma}&=&B^{\sigma}\nonumber\\
sB^{\tau}&=&0.
\end{eqnarray}
The dimensions of the fields are
$$
[g_{\mu \nu}]=0, \qquad [C^{\rho}]=0, \qquad [\bar{C}^{\sigma}]=0,\qquad [B^{\tau}]=1.
$$
It is easy to show that the operator $s$ is nilpotent, i.e. $s^2=0$. Therefore, as in the case of gauge theories, we can add to the classical action a gauge-fixing term that is $s$-exact, obtaining the gauge-fixed action
\begin{equation}
\label{stellegf}
S_{\text{gf}}=S_{\text{HD}}+s\Psi,
\end{equation}
where $\Psi$ is a fermionic functional with $[\Psi]=-1$. For the purposes of this section, we choose the functional
\begin{equation}\label{stellepsi}
\Psi=\alpha\int\bar{C}^{\mu }\square\left( \mathcal{G}_{\mu }-\lambda\kappa^{2}B_{\mu
}\right),
\end{equation}
where $\square=\eta^{\mu\nu}\partial_{\mu}\partial_{\nu}$, $\lambda$ is a gauge-fixing parameter and 
$\mathcal{G}_{\mu}$ is the gauge-fixing function. We choose
\begin{equation}
    \mathcal{G}_{\mu}=\eta ^{\nu \rho }\partial _{\rho }g_{\mu \nu },
\end{equation}
to have a simplified propagator. More general fermionic functionals and gauge-fixing function can help in checking gauge independence of the counterterms (see~\cite{Anselmi:2018ibi}). With these choices the gauge-fixing term reads
\begin{equation}
    s\Psi=\alpha\int B^{\mu }
\square \left( \mathcal{G}_{\mu }-\lambda\kappa^{2}B_{\mu
}\right) +S_{\text{gh}},  \label{skpsi}
\end{equation}%
where $S_{\text{gh}}$ is the action of the Faddeev-Popov ghosts
\begin{equation}
S_{\text{gh}}=\alpha\int \bar{C}^{\mu }\partial ^{\nu
}\square\left( g_{\mu \rho }\partial _{\nu
}C^{\rho }+g_{\nu \rho }\partial _{\mu }C^{\rho }+C^{\rho }\partial _{\rho
}g_{\mu \nu }\right) .  \label{eq:ghac}
\end{equation}
In order to obtain the graviton propagator we substitute $B$ with the solution of its equations of motion, i.e.
\begin{equation}
B_{\mu }=\frac{1}{2\lambda\kappa^{2}}\mathcal{G}_{\mu},
\end{equation}
and the gauge-fixed action \eqref{stellegf} becomes
\begin{equation}
S_{\text{gf}}=S_{\text{HD}}+\frac{\alpha}{4\lambda\kappa ^{2}}\int 
\mathcal{G}^{\mu } \square\mathcal{G}_{\mu }+S_{%
\text{gh}}.
\end{equation}
We write a generic quadratic part of the graviton action as
\begin{equation}\label{eq:gfhd}
    S_{\text{gf}}^{\text{quad}}=\int h_{\mu\nu}T^{\mu\nu\rho\sigma}h_{\rho\sigma},
\end{equation}
so the propagator $D_{\mu\nu\rho\sigma}$ is the Fourier transform of the Green function of $T^{\mu\nu\rho\sigma}$, i.e.
\begin{equation}\label{eq:propeq}
    \tilde{T}^{\mu\nu\alpha\beta}D_{\alpha\beta\rho\sigma}=\frac{i}{2}\left(\delta^{\mu}_{\rho}\delta^{\nu}_{\sigma}+\delta^{\mu}_{\sigma}\delta^{\nu}_{\rho}\right),
\end{equation}
where $\tilde{T}$ is the Fourier transform of $T$. Considering the symmetries of the indices, a basis for both $\tilde{T}$ and $D$ is given by the spin-2 projectors $\{\Pi^{(2)},\Pi^{(1)},\Pi^{(0)},\bar{\Pi}^{(0)}\}$ plus the tensor $\bar{\bar{\Pi}}^{(0)}$ that takes into account the possible mixing between two different spin-0 component, which are defined as follows
\begin{eqnarray}
\mbox{{\large
$\Pi $}}_{\mu \nu \rho \sigma }^{(2)}&\equiv&\frac{1}{2}(\pi_{\mu\rho}\pi_{\nu\sigma}+\pi_{\mu\sigma}\pi_{\nu\rho})-\frac{1}{3}\pi_{\mu\nu}\pi_{\rho\sigma},\\
\mbox{{\large
$\Pi $}}_{\mu \nu \rho \sigma }^{(1)}&\equiv&\frac{1}{2}(\pi_{\mu\rho}\omega_{\nu\sigma}+\pi_{\mu\sigma}\omega_{\nu\rho}+\pi_{\nu\rho}\omega_{\mu\sigma}+\pi_{\nu\sigma}\omega_{\mu\rho}),\\
\mbox{{\large
$\Pi $}}_{\mu \nu \rho \sigma }^{(0)}&\equiv&\frac{1}{3}\pi_{\mu\nu}\pi_{\rho\sigma},\\
\mbox{{\large
$\bar{\Pi} $}}_{\mu \nu \rho \sigma }^{(0)}&\equiv&\omega_{\mu\nu}\omega_{\rho\sigma},\\
\mbox{{\large
$\bar{\bar{\Pi}} $}}_{\mu \nu \rho \sigma }^{(0)}&\equiv&\pi_{\mu\nu}\omega_{\rho\sigma}+\pi_{\rho\sigma}\omega_{\mu\nu},
\end{eqnarray}
where $\pi_{\mu\nu}$ and $\omega_{\mu\nu}$ are the spin-1 projectors
\begin{equation}
\pi_{\mu\nu}\equiv\eta_{\mu\nu}-\frac{p_{\mu}p_{\nu}}{p^2}, \qquad\omega_{\mu\nu}\equiv\frac{p_{\mu}p_{\nu}}{p^2}.
\end{equation}
Therefore, we can write (omitting the indices)
\begin{equation}
\tilde{T}=x_2\Pi^{(2)}+x_1\Pi^{(1)}+x_0\Pi^{(0)}+\bar{x}_0\bar{\Pi}^{(0)}+\bar{\bar{x}}_0\bar{\bar{\Pi}}^{(0)}
\end{equation}
\begin{equation}
D=y_2\Pi^{(2)}+y_1\Pi^{(1)}+y_0\Pi^{(0)}+\bar{y}_0\bar{\Pi}^{(0)}+\bar{\bar{y}}_0\bar{\bar{\Pi}}^{(0)},
\end{equation}
where $x_i$ are obtained from the quadratic part of the action and $y_i$ are unknowns to be derived by imposing~\eqref{eq:propeq}. For a general theory of gravity, the relations between $x_i$ and $y_i$ are
\begin{equation}\label{eq:xy}
    y_1=\frac{i}{x_1}, \quad y_2=\frac{i}{x_2},\quad y_0=\frac{i\bar{x}_0}{x_0\bar{x}_0-3\bar{\bar{x}}_0^2},\quad \bar{y}_0=\frac{ix_0}{x_0\bar{x}_0-3\bar{\bar{x}}_0^2},\quad \bar{\bar{y}}_0=-\frac{i\bar{\bar{x}}_0}{x_0\bar{x}_0-3\bar{\bar{x}}_0^2}.
\end{equation}
After applying~\eqref{eq:xy} to the action~\eqref{eq:gfhd} and setting $\Lambda_C$ to zero for simplicity, the graviton propagator reads
\begin{equation}\label{stelleprop}
D_{\mu\nu\rho\sigma}=\frac{i }{p^2+i\epsilon}\left[\frac{\mbox{{\large
$\Pi $}}_{\mu \nu \rho \sigma }^{(2)}}{(\zeta-\alpha p^2)}-\frac{\mbox{{\large
$\Pi $}}_{\mu \nu \rho \sigma }^{(0)}}{2(\zeta-\xi p^2)}-\frac{\lambda}{2\alpha p^2}\left(2\mbox{{\large
$\Pi $}}_{\mu \nu \rho \sigma }^{(1)}+\mbox{{\large
$\bar{\Pi} $}}_{\mu \nu \rho \sigma }^{(0)}\right)\right].
\end{equation}
The expression of the Faddeev-Popov ghost propagator follows straightforwardly from the part of $S_{\text{gh}}$ that is quadratic in the fields $C$, $\bar{C}$. In momentum space we find
\begin{equation}\label{ghostpropstelle}
\left<C^{\mu}\bar{C}^{\nu}\right>=\frac{i}{p^2+i\epsilon}\left(\frac{\eta^{\mu\nu}-p^{\mu}p^{\nu}/2p^2}{\alpha p^2}\right).
\end{equation}
We identify the single poles and their residues by splitting the propagator \eqref{stelleprop} into partial fractions. To further simplify, we choose $\lambda=0$, then we find
\begin{equation}\label{eq:propstelle}
D_{\mu\nu\rho\sigma}^{\lambda=0}=\frac{i}{2\zeta}\left[\frac{2\mbox{{\large
$\Pi $}}_{\mu \nu \rho \sigma }^{(2)}-\mbox{{\large
$\Pi $}}_{\mu \nu \rho \sigma }^{(0)}}{p^2+i\epsilon}-\frac{2\mbox{{\large
$\Pi $}}^{(2)}_{\mu\nu\rho	\sigma}}{p^2-\zeta/\alpha+i\epsilon}+\frac{\mbox{{\large
$\Pi $}}_{\mu \nu \rho \sigma }^{(0)}}{p^2-\zeta/\xi+i\epsilon}\right].
\end{equation}
The first term is the same as in Einstein gravity and describes a massless graviton, while the second and third term are typical of Stelle gravity and describe a massive spin-2 field and a massive scalar field, respectively. The residue at the massive spin-2 pole is negative. This means that the theory propagates a ghost, which violates unitarity. In order to make this clear, we show a simple example by means of a scalar toy model. Consider the action
\begin{equation}\label{eq:phi3hd}
    S(\varphi)=\int\left[\frac{1}{2}\partial_{\mu}\varphi\left(1+\frac{\square}{M^2}\right)\partial^{\mu}\varphi-\frac{1}{2}m^2\varphi\left(1+\frac{\square}{M^2}\right)\varphi-\frac{\lambda}{3!}\varphi^3\right],
\end{equation}
where $\varphi$ is a scalar field, $m$ and $M$ are mass parameters and $\lambda$ is a coupling constant. Using the Feynman prescription, the propagator reads
\begin{equation}\label{eq:propscalar4}
    D(p^2)=\frac{-iM^2}{(p^2-m^2+i\epsilon)(p^2-M^2+i\epsilon)}=\frac{i}{p^2-m^2+i\epsilon}-\frac{i}{p^2-M^2+i\epsilon},
\end{equation}
where in the second step we have appropriately redefined $\epsilon$. A degree of freedom associated to a propagator with negative residue at the pole is called ghost. The propagator~\eqref{eq:propscalar4} describes two degrees of freedom, one at $p^2=m^2$ with a positive residue and one at $p^2=M^2$ with negative residue. Ghosts should not be confused with tachyons, which are defined as degrees of freedom with negative mass squared\footnote{Note that a ghost can also be a tachyon.} and introduce additional pathologies. For the present review it is important just to know that tachyons cannot be cured by the fakeon prescription and the parameters of the theory must be constrained in order to avoid tachyons (see~\autoref{sec:QGcosm}).

The simplest example of a unitarity violation is given by the tree-level 2-to-2 scattering amplitude. Unitarity is encoded in the optical theorem (see next section for the details), which, in the case the process considered, states that
\begin{equation}\label{eq:opt}
    2\text{Im}\mathcal{M}_{2\rightarrow 2}=\int\mathrm{d}\Pi\left|\mathcal{M}_{2\rightarrow 1}\right|^2\geq 0,
\end{equation}
where $\mathcal{M}_{2\rightarrow 2}$ and $\mathcal{M}_{2\rightarrow 1}$ denote the tree-level amplitudes of the processes $\varphi\varphi\rightarrow\varphi\varphi$ and $\varphi\varphi\rightarrow\varphi$, respectively, while $\text{d}\Pi$ is the integration measure over the phase space of final states. The amplitude $\mathcal{M}_{2\rightarrow 2}$ is
\begin{equation}
    \mathcal{M}_{2\rightarrow 2}=-\lambda^2\left(\frac{1}{p^2-m^2+i\epsilon}-\frac{1}{p^2-M^2+i\epsilon}\right)
\end{equation}
and its imaginary part reads
\begin{equation}
  \text{Im}\mathcal{M}_{2\rightarrow 2}=\pi\lambda^2\left[\delta(p^2-m^2)-\delta(p^2-M^2)\right], 
\end{equation}
which is not nonnegative and therefore violates~\eqref{eq:opt}. Similar violations appear every time a odd number of ghosts is on shell in the right-hand side of~\eqref{eq:opt}. 

This simple example shows what happens also in Stelle theory. Moreover, note that the pole of the additional scalar in~\eqref{eq:propstelle} has positive residue. Therefore, if we send the mass of the ghost to infinity ($\alpha\rightarrow 0$) we find a unitary theory describing a massless graviton and a massive scalar, the Starobinsky model. However, in that case renormalizability is lost. In fact, the presence of the ghost is crucial to obtain certain cancellations between Feynman diagrams that avoid higher-dimensional counterterms to be generated by renormalization. In the end renormalizability can be obtained at the price of unitarity. These two properties seem to be mutually exclusive in quantum gravity, since they are both entangled to the ghost. If we have the ghost we loose unitarity, while if we get rid of the ghost we loose renormalizability. To define a quantum field theory of gravity that makes sense both physically and mathematically we need to find a way to consistently remove the ghost from the set of physical states and keep its contributions inside Feynman diagrams at the same time. This cannot be achieved without relaxing some assumptions of standard quantum field theory. In our approach we disentangle the on-shell contributions of the ghost from the virtual ones, that would typically be related in the standard case of the Feynman prescription. Once this modification is introduced we can obtain the property we want and remove the ghost from the spectrum at any energy scale without spoiling the consistency of the theory. 

Before to proceed with the details of the fakeon prescription we give a sketch of the proof of renormalizability in Stelle gravity for the readers that are not familiar with it. We anticipate that the fakeon prescription do not modify the divergences of the theory, therefore from this point of view there is no difference with the Stelle gravity.

\subsection{Renormalization}
We give the main ingredients to prove the renormalizability of the theory with the help of the Batalin-Vilkovisky formalism, which we briefly introduce below. In~\cite{Stelle:1976gc} it was assumed that the divergent part of the effective action satisfies the so-called Kluberg-Stern--Zuber conjecture~\cite{Kluberg-Stern:1974nmx,Kluberg-Stern:1975ebk}. More recently it has been proved by a theorem in~\cite{Anselmi:2015niw} for a large class of models, which includes higher-derivative quantum gravity. We do not review the proof here and the reader is referred to~\cite{Anselmi:2015niw} for details.

The Batalin-Vislkovisky formalism is a useful tool to prove renormalizability in gauge theories and gravity. It makes use of an extended action $\Sigma$ that accounts for the sources of the BRST transformations and allows to write the Ward-Takahashi-Slavnov-Taylor identities in a compact form. The counterterms then satisfiy certain properties and the divergences are removed by means of an iterative procedure. 

We collect all the fields into the row $\Phi^{\alpha}=(g_{\mu\nu},C^{\rho},\bar{C}^{\sigma},B^{\tau})$ and introduce a set of sources $K_{\alpha}=(K_g^{\mu\nu},K^{C}_{\sigma},K_{\bar{C}}^{\tau},K_B^{\tau})$ for the composite BRST operators $s\Phi^{\alpha}$, conjugate to the fields, and define the \textit{antiparentheses} of two
functionals $X$ and $Y$ of $\Phi $ and $K$ as 
\begin{equation}
(X,Y)\equiv \int \left( \frac{\delta _{r}X}{\delta \Phi ^{\alpha }}\frac{%
\delta _{l}Y}{\delta K_{\alpha }}-\frac{\delta _{r}X}{\delta K_{\alpha }}%
\frac{\delta _{l}Y}{\delta \Phi ^{\alpha }}\right) ,
\end{equation}%
where the integral is over the spacetime points associated with repeated
indices and the subscripts $l$, $r$ in $\delta _{l}$, $\delta _{r}$ denote
the left and right functional derivatives, respectively. We introduce the \emph{ghost number} 
\begin{equation}
    \text{gh}\#(g_{\mu\nu})=\text{gh}\#(B^{\tau})=0, \qquad \text{gh}\#(C^{\rho})=1, \qquad \text{gh}\#(\bar{C}^{\sigma})=-1,
\end{equation}
so that the gauge-fixed action is invariant under the global symmetry
\begin{equation}\label{eq:ghostsymm}
    \Phi^{\alpha}\rightarrow\Phi^{\alpha}e^{i\beta \text{gh}\#(\Phi^{\alpha})},
\end{equation}
where $\beta$ is a constant. 
Finally, the extended action is defined by adding the source term $S_K=-\int s\Phi^{\alpha}K_{\alpha}$ to the gauge-fixed action. The dimensions of the sources are
\begin{equation}
[K_{g}]=3,\quad [K_C]=3,\quad [K_{\bar{C}}]=3,\quad [K_B]=2,
\end{equation}
while the ghost numbers are fixed by requiring that $S_K$ be invariant under the symmetry~\eqref{eq:ghostsymm} combined with\footnote{This symmetry is not spoiled by radiative corrections since both Feynman rules and the diagrammatics are compatible with it.}
\begin{equation}
    K_{\alpha}\rightarrow K_{\alpha}e^{i\beta \text{gh}\#(K_{\alpha})}
\end{equation}
and read
\begin{equation}
\text{gh}\#(K_{g})=-1,\quad \text{gh}\#(K_C)=-2,\quad \text{gh}\#(K_{\bar{C}})=0,\quad \text{gh}\#(K_B)=-1.
\end{equation}
Finally, the statistics of the fields $\Phi^{\alpha}$ and sources $K_{\alpha}$ are opposite to each other.

In the Batalin-Vilkovisky formalism, the gauge-fixing term can be written as
\begin{equation}
    s\Psi=(S_K,\Psi)
\end{equation}
and the extended action reads
\begin{equation}\label{eq:sigmastelle}
    \Sigma(\Phi,K)=S_{\text{HD}}+(S_K,\Psi)+S_K.
\end{equation}
This action is used to define the generating functional $Z$ of the correlation functions and the generating functional $W$ of the connected correlation functions by means of the formula
\begin{equation}
   Z(J,K)=\int [\mathrm{d}\Phi ]\exp \left( i\Sigma(\Phi ,K)+i\int \Phi ^{\alpha
}J_{\alpha }\right) =\exp iW(J,K).
\end{equation}
As usual, the generating functional of one-particle irreducible diagrams is defined as the Legendre transform of $W$ with respect to $J$
\begin{equation}
    \Gamma(\Phi,K)=W(J,K)-\int\Phi^{\alpha}J_{\alpha},
\end{equation}
where $\Phi^{\alpha}=\delta_rW/\delta J_{\alpha}$. Renormalization is achieved by means of parameter and field redefinitions, which are canonical transformations with respect to the antiparentheses. It is easy to show that $\Sigma$ and $\Gamma$ satisfy the \textit{master equations}
\begin{equation}
(\Sigma,\Sigma)=0, \qquad \mbox{and}\qquad (\Gamma,\Gamma)=0.
\end{equation}
The first one is a generalization of the BRST invariance of the gauge-fixed action. In fact, the operator
\begin{equation}
    \sigma\equiv(\Sigma,\cdot)
\end{equation}
reduces to the BRST transformations when acting on $\Phi$, while it further reduces to a diffeomorphism when acting on $g_{\mu\nu}$. The second equation collects the Ward-Takahashi-Slavnov-Taylor identities. Moreover, it contains information about the counterterms. In order to keep track of the loop expansion, we temporarily reintroduce the factors $\hbar$, the $n$-th order in $\hbar$ being the $n+1$-th order in loops. Then, by expanding $\Gamma$ in powers of $\hbar$ and using the master equation, it is possible to show that the divergent part of the effective action satisfies the equation
\begin{equation}\label{eq:gammadivstelle}
\sigma\Gamma_{\mathrm{div}}^{(n)}=0,
\end{equation}
where $\Gamma_{\mathrm{div}}^{(n)}$ is the divergent part of the functional $\Gamma^{(n)}$, which we assume to be convergent up to the order $\hbar^{n-1}$. If a functional $X$ satisfies $\sigma X=0$, we say that it is \textit{$\sigma$-closed}, while if it is such that $X=\sigma Y$, where $Y$ is another functional, we say that it is \textit{$\sigma$-exact}. The general solution of~\eqref{eq:gammadivstelle} can be written as
\begin{equation}\label{eq:solutionstelle}
\Gamma_{\mathrm{div}}^{(n)}(\Phi,K)=\tilde{G}_{n}(\Phi,K)+\sigma\tilde{X}_{n}(\Phi,K),
\end{equation}
where $\tilde{G}_{n}$ is a $\sigma$-closed local functional. In order to prove renormalizability we need to ensure that it is possible to move all the dependence on the sources and unphysical fields into the $\sigma$-exact term. This means that the solution~\eqref{eq:solutionstelle} is reorganized in the form
\begin{equation}\label{eq:gammadivstellered}
\Gamma_{\mathrm{div}}^{(n)}(\Phi,K)=G_{n}(h)+\sigma X_{n}(\Phi,K),
\end{equation}
where the functional $G_{n}$ now depends only on the metric fluctuation and is gauge invariant, being $\sigma$-closed. This is the Kluberg-Stern--Zuber conjecture mentioned above, which is now proved by a theorem~\cite{Anselmi:2015niw}. We give some additional properties of $\Sigma$, $\Gamma_{\text{div}}^{(n)}$ and $X_n$, which allow us to write $\Gamma_{\text{div}}^{(n)}$ in a more explicit form
\begin{itemize}
\item[\textbf{1)}] The divergent part cannot depend on $B$, $K_{\bar{C}}$ and $K_B$ because no vertices of the action~\eqref{eq:sigmastelle} contain them, so no one-particle irreducible diagrams can be built with $B$, $K_{\bar{C}}$ and $K_B$. 

\item[\textbf{2)}]$\Sigma$ depends on $K_{g}$ and $\bar{C}$ only through the combination 
\begin{equation}\label{Kdepstelle}
\tilde{K}_{g}^{\mu\nu}\equiv K_{g}^{\mu\nu}+\alpha\Box\int\frac{\delta\mathcal{G}_{\rho}}{\delta g_{\mu\nu}}\bar{C}^{\rho}.
\end{equation}
This property ensures that also $\Gamma_{\text{div}}^{(n)}$ depends on $K_g$ and $\bar{C}$ only through $\tilde{K}_g$. In fact, given a diagram with an external $K_g^{\mu\nu}$ leg, there exists an almost identical diagram where the $K_g^{\mu\nu}$ leg is replaced by a $\alpha\Box\int\frac{\delta\mathcal{G}_{\rho}}{\delta g_{\mu\nu}}\bar{C}^{\rho}$ leg and viceversa.
\item[\textbf{3)}] $X_n$ also depends on $K^{\mu \nu }_g$ and $\bar{C}^{\rho }$ via the combination $\tilde{K}^{\mu \nu}_g$. The proof of this property is as follows. The dimension of $X_n$ and its ghost
number imply that we can parametrize it as 
\begin{equation}\label{fpara}
X_n(\Phi,K)=\int\Delta g_{\mu\nu}K_g^{\mu\nu}+\int\Delta C^{\rho}K^{C}_{\rho}+\int\bar{C}^{\rho}L_{\rho},
\end{equation}
where $L_{\rho}$ is a function of dimension $3$ and ghost number zero, while $\Delta g_{\mu \nu }$ and $\Delta C^{\rho }$ are the renormalizations
of the metric tensor and the Faddeev-Popov ghosts, respectively. Then, from the expression \eqref{eq:gammadivstellered} and property 1) it follows that also $\sigma X_n$ does not depend on $B$. In $\sigma X_n$ the terms linear in $B$ are
\begin{equation}\label{sfb}
\left.\sigma X_n\right|_{B}=\int\left.\left(\frac{\delta_r S}{\delta g_{\mu\nu}}\frac{\delta_l X_n}{\delta K_g^{\mu\nu}}-\frac{\delta_r X_n}{\delta \bar{C}^{\rho}}\frac{\delta_l S}{\delta K^{\bar{C}}_{\rho}}\right)\right|_B=\int B^{\rho}\left[\alpha
\square\int\frac{\delta_r\mathcal{G}_{\rho }}{\delta g_{\mu\nu}}\Delta g_{\mu\nu}+L_{\rho}\right].
\end{equation}
Therefore, the expression in the squared bracket in \eqref{sfb} must vanish, which implies that the terms proportional to $K_g$ plus those proportional to $\bar{C}$ in $X_n$ are 
\begin{eqnarray}
\left. X_n(\Phi ,K)\right|_{K_g,\bar{C}}&=&\int\left[ \Delta g_{\mu\nu}K_g^{\mu \nu }-\bar{C}^{\rho}\alpha
\square\int\frac{\delta_r\mathcal{G}_{\rho }}{\delta g_{\mu\nu}}\Delta g_{\mu\nu}\right]\nonumber\\
&=&\int\Delta g_{\mu\nu}\tilde{K}_g^{\mu \nu }.
\end{eqnarray}
\end{itemize}
From the properties listed above we deduced that \begin{equation}
    X_n(\Phi,K)=\int\Delta g_{\mu\nu}\tilde{K}_g^{\mu\nu}+\int\Delta C^{\rho}K^{C}_{\rho}.
\end{equation}
Although $\Gamma_{\text{div}}^{(n)}$ can be written in the form~\eqref{eq:gammadivstellered}, it is still possible that the $\sigma$-exact term contains terms that depend only on the physical fields. More explicitly, a straightforward calculation gives%
\begin{equation}
\sigma X_n=\int \frac{\delta S_{\text{HD}}}{\delta g_{\mu \nu }}\Delta
g_{\mu \nu }-\int \Delta \mathcal{R}_{\mu \nu }\tilde{K}^{\mu \nu }-\int
\Delta \mathcal{R}^{\rho }K_{\rho }^{C},  \label{sf}
\end{equation}%
where%
\begin{eqnarray}
\Delta \mathcal{R}_{\mu \nu } &=&-\sigma\Delta g_{\mu \nu }+\int \Delta
g_{\alpha \beta }\frac{\delta _{l}(sg_{\mu \nu })}{\delta
g_{\alpha \beta }}+\int \Delta C^{\tau }\frac{\delta _{l}(sg_{\mu
\nu })}{\delta C^{\tau }}, \\[2ex]
\Delta \mathcal{R}^{\rho } &=&-\sigma\Delta C^{\rho }+\int \Delta C^{\tau }%
\frac{\delta _{l}(sC^{\rho })}{\delta C^{\tau }}
\end{eqnarray}%
We can see that there is a term proportional to the equations of motion, which depends only on the metric. Then, if we split the metric as in~\eqref{eq:metricsplit}  we have 
\begin{equation}
\Delta g_{\mu\nu}=t_0g_{\mu\nu}+t_1 h_{\mu\nu}+t_2 \eta_{\mu\nu}h+\mathcal{O}(h^2),
\end{equation}
where $t_i$ are (gauge-dependent) coefficients that depend on the parameters in~\eqref{lhd}, which makes the term proportional to the equations of motion non covariant. Therefore, when we compute the divergent part of a Feynman diagrams with only graviton external legs, we would get also the terms proportional to the equations of motion. In practice, this mean that those terms always need to be taken into account when computing the divergent part of diagrams in Stelle theory and, in general, the results obtained in this way are not covariant. In other words, the functional $G_n$ cannot be written as a function of $g_{\mu\nu}$ only but it depends also on $h_{\mu\nu}$ in an independent way. Only after removing those terms, we can safely conclude that any other one-particle irreducible diagrams would generate divergent terms that sum together to give a covariant result. Another way to see this is to consider the wave functions renormalization of $h_{\mu\nu}$ and its trace as independent. This depends on the gauge choice since the wave function renormalization are unphysical. For example, the background field method, which can be seen as a gauge choice, takes care of terms like those in $\sigma X_n$. If we do not choose a particular gauge or a method that preserves general covariance at each steps, we can remove the $\sigma$-exact term by means of the transformation 
\begin{equation}\label{eq:canonical}
{\Phi^i}^{\prime}=\Phi^i-\frac{\delta X_{n}(\Phi,K)}{\delta K_{i}^{\prime}}, \qquad K_{i}^{\prime}=K_i-\frac{\delta X_{n}(\Phi,K)}{\delta \Phi^i}.
\end{equation}
which is canonical, i.e. preserves the antiparentheses. In components, it reads
\begin{equation}
{h^{\prime}}_{\mu\nu}=h_{\mu\nu}-\Delta g_{\mu\nu},\qquad
{C^{\prime}}^{\sigma}=C^{\rho}-\Delta C^{\rho},\qquad
\bar{C^{\prime}}^{\tau}=\bar{C}^{\tau},\qquad
{B^{\prime}}^{\nu}=B^{\nu},
\end{equation}
while the sources transform as 
\begin{equation}
\begin{split}
K^{\prime\mu\nu}_g=K^{\mu\nu}_g+\tilde{K}^{\alpha\beta}_g&\frac{\delta\Delta g_{\alpha\beta}}{\delta h_{\mu\nu}}-K^{C}_{\sigma}\frac{\delta \Delta C^{\sigma}}{\delta h_{\mu\nu}}, \qquad
K^{\prime C}_{\sigma}=K^{C}_{\sigma}-K^{C}_{\alpha}\frac{\delta \Delta C^{\alpha}}{\delta C^{\sigma}},\\[2ex]
&K^{\prime \bar{C}}_{\tau}=K^{\bar{C}}_{\tau},\qquad
K^{\prime B}_{\tau}=K^{B}_{\tau}.
\end{split}
\end{equation}
We obtain
\begin{equation}
\Gamma_{\text{div}}^{(n)}(\Phi^{\prime},K^{\prime})=\Gamma_{\text{div}}^{(n)}(\Phi,K)-\sigma X_{n}(\Phi, K)=\int 
\sqrt{-g}\left[ 2\Lambda_{n}+\zeta_{n} R+\frac{\alpha_{n}}{2}C^2 -\frac{\xi_{n} }{6}R^{2}%
\right],
\end{equation}
where all the terms of dimension 0, 2 and 4 have been included as possible counterterms and the parameters with the subscript $n$ denote the renormalization of the associated quantities.

We stress that we used the fact that the divergent terms can be organized as~\eqref{eq:gammadivstellered} in any gauge. Additional details can be found in~\cite{Anselmi:2015niw} and~\cite{Anselmi:2018ibi}. However, we could choose a gauge fixing and a parametrization of the graviton field~\cite{Stelle:1976gc} such that $\Delta g_{\mu\nu}=\Delta C^{\rho}=0$, so  $X_n$ is zero and the transformation~\eqref{eq:canonical} is trivial. In that case, no terms proportional to the equations of motion appear, the counterterms are gauge invariant and there is no need to use the theorem that shows~\eqref{eq:gammadivstellered} to prove renormalizability in that specific setting.

\sect{Unitarity and cutting identities}\label{sec:unit}
In this section we recall how unitarity is formulated diagrammatically by using the so-called \textit{cutting equations}. Unitarity is the condition
\begin{equation}
    SS^{\dagger}=\mymathbb{1}
\end{equation}
on the $S$ matrix, which can be rewritten in terms of its nontrivial part $T$ as
\begin{equation}\label{eq:optth}
-i(T-T^{\dagger})=TT^{\dagger}.
\end{equation}
The matrix elements of equation~\eqref{eq:optth} are obtained by choosing initial and final states, $|i\rangle$, $|f\rangle$ and by inserting the completeness relation between $T$ and $T^{\dagger}$ on the right-hand side of~\eqref{eq:optth}
\begin{equation}\label{eq:optthme}
   -i\langle f|(T-T^{\dagger})|i\rangle=\sum_{|n\rangle}\int \mathrm{d}\Pi_n\langle f|T|n\rangle\langle n|T^{\dagger})|i\rangle,
\end{equation}
where the sum runs over all the possible final states and the integral symbolically denote the integration over the phase space of those final states. The matrix element $\langle f|T|i \rangle$ is associated to the amplitude $\mathcal{A}_{fi}$ up to a $\delta$ function for the conservation of four momentum. If we write the equation~\eqref{eq:optth} in terms of the amplitudes for an elastic scattering, i.e. $|f\rangle=|i\rangle$, we obtain
\begin{equation}
    2\text{Im}\mathcal{A}_{ii}^{(\text{f})}(s)=\sum_{n}\int\mathrm{d}\Pi_n|\mathcal{A}_{in}|^2=2\Phi \sigma_{\text{tot}}(s),
\end{equation}
where $\mathcal{A}^{(\text{f})}$ is the forward amplitude, $s$ is the center-of-mass energy squared, $\Phi$ is the flux factor of initial particles and $\sigma_{\text{tot}}$ is the total cross sections. This has a more familiar form known from the optical theorem in nonrelativistic scattering and this is why also~\eqref{eq:optth} is named like that. However, equation~\eqref{eq:optth} is more general, since it is valid for any type of process and on its right-hand side typically involves quantities that are not cross sections. Since quantum field theory is defined perturbatively by means of Feynman diagrams, equation~\eqref{eq:optth} can be expanded diagrammatically, obtaining a set of equations order by order. There is a more powerful set of identities that are valid diagram by diagram: the cutting equations, which hold in any local quantum field theory\footnote{Here locality means that it is possible to build consistent Feynman rules that involve local vertices.}. From now on we refer to them as \textit{cutting identities}. However, only if the theory satisfies additional requirements the cutting identities imply the optical theorem.

The first step to derive the cutting identities is the \textit{largest-time equation}, which is their coordinate-space version. For the purpose of this section we consider only the scalar $\varphi^3$ theory given by the action~\eqref{eq:phi3hd} without the higher-derivative terms. The generalization to fermions, gauge theories and gravity involves some caveats related to gauge invariance, which we do not address here. More details can be found in~\cite{tHooft:1973bhk, Anselmi:2016fid}. First, we recall that the Feynman propagator 
\begin{equation}
D_{ij}(x)=\int\frac{\mathrm{d}^4p}{(2 \pi)^4}\frac{i}{p^2-m^2+i\epsilon}e^{-ipx}, \qquad x\equiv x_i-x_j
\end{equation}
can be written as
\begin{equation}\label{eq:feyndec}
    D_{ij}(x)=\theta(x^0)D_{ij}^+(x)+\theta(-x^0)D_{ij}^-(x),
\end{equation}
where
$$
D_{ij}^{\pm}(x)=\int \frac{\mathrm{d}^3\mathbf{p}}{(2\pi)^3}\frac{e^{\mp i\omega(\mathbf{p})x^0}e^{\pm i \mathbf{p}\cdot\mathbf{x}}}{2\omega(\mathbf{p})}, \qquad \omega(\mathbf{p})\equiv \sqrt{\mathbf{p}^2+m^2},
$$ 
$\theta$ is the Heaviside step function and the bold symbols represent the space components of vectors. Then, consider a generic Feynman diagram represented by a function $F(x_1,\ldots,x_n)$, where each spacetime point $x_i$ is associated to a vertex. For example, a one-loop three-point function is given by
\begin{equation}
    F(x_1,x_2,x_3)=(-i\lambda)^3D_{12}D_{23}D_{31},
\end{equation}
where $\lambda$ indicates a generic coupling constant. We introduce another function where one or more $x_i$ are marked by a hat. The new function is obtained from $F$ by means of the following substitutions
\begin{equation}
    D_{ij}(\hat{x}_i-x_j)\rightarrow D_{ij}^+, \qquad D_{ij}(x_i-\hat{x}_j)\rightarrow D_{ij}^-, \qquad D_{ij}(\hat{x}_i-\hat{x}_j)\rightarrow (D_{ij})^*.
\end{equation}
Moreover, each vertex associated with a marked point must be substituted with its complex conjugate. Using the example above we can have
\begin{equation}
F(\hat{x}_1,\hat{x}_2,x_3)=(-i\lambda)^3(D_{12})^*D^+_{23}D^-_{31}.
\end{equation}
Now, suppose that a time component $x_i^0$ for some $i$ is larger than all the others. Then any diagram where $x_i$ is not marked is equal to minus the same diagram in which $x_i$ is marked. This follows straightforwardly from the definition of marked points and from~\eqref{eq:feyndec}. For example, in the case of the three-point function above, assuming that $x_1^0$ is the largest time component, we can write
\begin{equation}
F(x_1,x_2,x_3)=(-i\lambda)^3D_{12}D_{23}D_{31}=(-i\lambda)^3D^+_{12}D_{23}D^-_{31}=-F(\hat{x}_1,x_2,x_3),
\end{equation}
\begin{equation}
F(x_1,x_2,\hat{x}_3)=-(-i\lambda)^3D_{12}D^-_{23}D^+_{31}=-(-i\lambda)^3D^+_{12}D^-_{23}D^+_{31}=-F(\hat{x}_1,x_2,\hat{x}_3)\end{equation}
and so on. Therefore, for any time configuration we have a set of identities that hold by construction, as long as it is possible to determine which time component is the largest, i.e. the vertices are local, and they have distinct times. Coincident points can generate contact terms which do not spoil the result~\cite{Anselmi:2016fid}. We can write the largest-time equation in a compact form without specifying which time component is the largest as
\begin{equation}\label{eq:lte}
\sum_{m}F_m(x_1,\ldots,\hat{x}_i,\ldots,\hat{x}_j,\ldots,x_n)=0,
\end{equation}
where the sum runs over all the possible ways of marking the points (including the cases where all and no vertices are marked) and the $F_m$ are the diagram with marked vertices. Then we can take the Fourier transform of~\eqref{eq:lte}. The Fourier transform of $D^{\pm}$ can be written in the form
\begin{equation}\label{eq:Dpmtilde}
    \tilde{D}^{\pm}(p)=2\pi\theta(p^0)\rho(p^2),
\end{equation}
where $\rho(p^2)$ is a distribution such that the Fourier transform of the propagator is
\begin{equation}
    \tilde{D}(p)=\int_{0}^{\infty}\frac{\mathrm{d}s}{2 \pi}\frac{i\rho(s)}{p^2-s+i\epsilon}.
\end{equation}
\begin{figure}
\centering
\includegraphics[width=15cm]{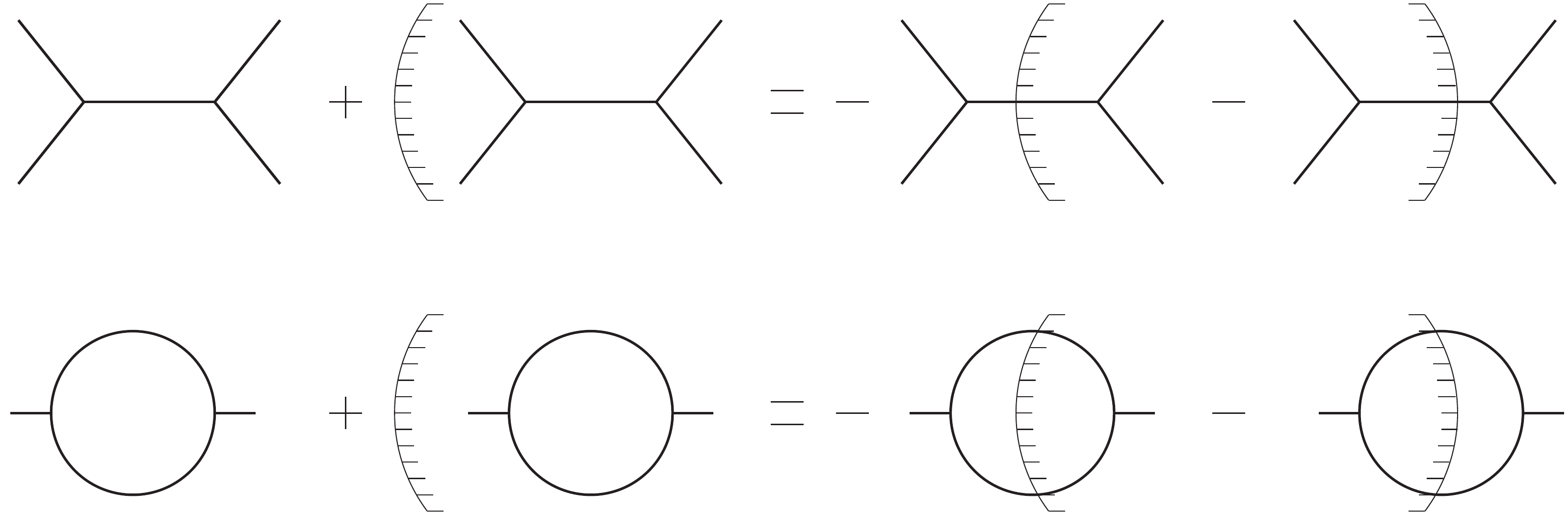}
\caption{On the top, the cutting identity for a tree-level diagram. On the bottom, the cutting identity for the bubble diagram.}
\label{fig:cutdiag}
\end{figure}

In the case of a standard scalar field theory we have
\begin{equation}
\rho(p^2)=\delta(p^2-m^2), \qquad \tilde{D}(p^2)=\frac{i}{p^2-m^2+i\epsilon}, \qquad \tilde{D}^{\pm}=2\pi\theta(\pm p^0)\delta(p^2-m^2).
\end{equation}
In order to have a better graphical view, we introduce cut diagrams, instead of marked ones, where an internal line is cut if it connects a marked vertex with an unmarked one. The cut is represented by a continuous line and has a ``shaded region" on the side where the marked vertex lies, which represents the energy flow given by the $\theta$-functions in~\eqref{eq:Dpmtilde}. In practice, the cut diagrams are obtained from the main diagram by means of the substitution
\begin{equation}
    \frac{i}{p^2-m^2+i\epsilon}\rightarrow 2\pi \theta(\pm p^0)\delta(p^2-m^2)
\end{equation}
for each cut propagator and by complex conjugating everything that lies on the shadowed region. Finally, a diagram where all the vertices lie in the shadowed region correspond to the complex conjugate of the diagram with no cut. With these definitions we can write the cutting identities
\begin{equation}\label{eq:cuttingid}
G+G^*=-\sum_{\text{cuts}}G_c,
\end{equation}
where $G$ is the diagram in momentum space and $G_c$ are the cut diagrams. Simple examples of the cutting identities are given by those in~\autoref{fig:cutdiag}. The one on the top reads
\begin{equation}\label{eq:cutidprop}
\begin{split}
-\lambda^2\left(\frac{i}{p^2-m^2+i\epsilon}+\frac{-i}{p^2-m^2-i\epsilon}\right)&=-\lambda^22\pi\theta(p^0)\delta(p^2-m^2)-\lambda^22\pi\theta(-p^0)\delta(p^2-m^2)\\
&=-\lambda^22\pi\delta(p^2-m^2).
\end{split}
\end{equation}
The identity on the bottom in~\autoref{fig:cutdiag} is shown by considering equal masses $m$ for the particles in the loop. In that case the bubble diagram reads
\begin{equation}\label{eq:bubblediag}
    \mathcal{B}(p^2)=-\frac{i\lambda^2}{2(4\pi)^2}\int_0^1\mathrm{d}x\ln \left(\frac{-xp^2(1-x)+m^2-i\epsilon}{\mu^2}\right)+\mathcal{B}_{\text{div}},
\end{equation}
where $x$ is a Feynman parameter, $\mu$ is the renormalization scale and $\mathcal{B}_{\text{div}}$ is the divergent part of the diagram evaluated in dimensional regularization, which might contain also some finite constant depending on the scheme. Therefore,
\begin{equation}\label{eq:imbub}
\mathcal{B}+\mathcal{B}^*=-\frac{\lambda^2}{16\pi}\theta(p^2-4m^2)\sqrt{1-\frac{4m^2}{p^2}}.
\end{equation}
Note that $\mathcal{B}_{\text{div}}$ cancels in~\eqref{eq:imbub}, since it is always purely imaginary. The cut diagrams are
\begin{equation}
\mathcal{B}_{12}=\lambda^2\int\frac{\mathrm{d}^4q}{(2\pi)^2}\theta(p^0+q^0)\delta((p+q)^2-m^2)\theta(-q^0)\delta(q^2-m^2)
\end{equation}
and analogous expression for $\mathcal{B}_{21}$, where the subscripts indicate the energy flow. It is easy to check the identity by direct computation and show that
\begin{equation}
\mathcal{B}+\mathcal{B}^*=-\mathcal{B}_{12}-\mathcal{B}_{21}.
\end{equation}
In this case, where the scalar field propagator has a positive residue, the cutting identities are equivalent to the optical theorem. For example, remembering that $G=iT$, the right-hand side of~\eqref{eq:imbub} is equal to the right-hand side of~\eqref{eq:optthme} when the initial state is a single particle with center-of-mass energy squared $p^2$ and the final state is two particles of mass $m^2$. Note that the $\theta$-function in~\eqref{eq:imbub} gives the correct kinematical condition $p^2>4m^2$ for the decay to happen.  However, it is not always true that~\eqref{eq:cuttingid} implies~\eqref{eq:optth}. In fact, we did not assume any positivity property on the distribution $\rho$ in~\eqref{eq:Dpmtilde} to prove~\eqref{eq:cuttingid}, which, instead, is crucial for the optical theorem. The simplest example is the case of a ghost. In that case we have $\rho=-\delta(p^2-m^2)$ and the right-hand side of~\eqref{eq:cutidprop} does not have the correct sign to match the right-hand side of~\eqref{eq:optthme}. 

In general we can say that the cutting identities implies a modified version of the optical theorem called \textit{pseudounitarity equation}, which, written in a compact form, reads
\begin{equation}
-i(T-T^{\dagger})=TCT^{\dagger}, \qquad C=\text{diag}(1,\ldots,1,-1,\ldots,-1,\ldots),
\end{equation}
where the minus ones in the matrix $C$ depend on the presence of ghosts in the theory. Therefore, the task of determine whether a theory is unitary or not reduces to find a way to deal with the matrix $C$. For example, in gauge theories the matrix $C$ is not the identity, due to the presence of Faddeev-Popov ghosts as well as longitudinal and temporal component of the gauge fields. However, thanks to the BRST symmetry it is possible to project the Fock space onto a subspace which is generated only by creation operators of the physical fields, i.e. the transverse gauge bosons. This operation is consistent with unitarity and the degrees of freedom that are projected out in this way are not generated back by loop corrections. In the next section we show that it is possible to achieve the same type of projection without the help of a symmetry by changing the quantization prescription for the degrees of freedom that we want to remove from the spectrum. However, there is a crucial difference in the case of the fakeon prescription: unlike the BRST case, purely virtual particles are not unphysical, so they can still contribute to the interactions between other standard degrees of freedom and give physical effects.

\sect{Spectral identities and unitarity with fakeons}
\label{sec:spectid}

In this section we show how to turn a standard degree of freedom into a purely virtual particle. This result is achieved by means of a set of operations that we call the \emph{fakeon prescription}. There are a few different equivalent ways of implementing the fakeon prescription. Here we present what we believe to be the clearest one, i.e. via the so-called \emph{threshold decomposition}. The amplitudes are obtained by using the Feynman prescription for every propagator in the theory and only at the end an additional step is performed. In this way it is clear which properties of the standard Feynman prescription survive the fakeon prescription and which get modified. Roughly speaking, what we want to achieve with the fakeon prescription is to remove all the parts of the amplitudes that are related to the  (would-be-purely-virtual) particle being on-shell. For example, the real part of the bubble diagram~\eqref{eq:bubblediag} tells whether the particle in the external line can decay into the (on-shell) particles in the loop, as shown in~\eqref{eq:imbub}. Once those pieces are subtracted, we perform a projection at the level of the Fock space by choosing to work in a subspace where the particles that we want to make purely virtual are not external legs. Combining the two operations (removing pieces of the amplitudes and projecting onto a subspace) guarantees that quantum corrections will not generate back the degrees of freedom that we remove from the set of particle that can appear on shell. 

In order to identify the subtractions, we decompose the amplitudes as sums of terms that are associated to single thresholds. In Feynman diagrams thresholds tell which kinematic configurations allow for the virtual particles to become on shell. To obtain the threshold decomposition it is sufficient to integrate each Feynman diagram over the loop energies. The integral over space momenta is postponed to after the decomposition is performed. 

We consider one-loop diagrams, the generalization to higher loops can be found in~\cite{Anselmi:2021hab}. Moreover, we derive the decomposition for a scalar theory, since all the crucial points are related to the singularities and branch cuts of the amplitudes, which are not modified by the presence of nontrivial numerators. For a one-loop diagram with $N$ legs, we define the \emph{skeleton diagram} as 
\begin{equation}
G_{N}^{s}=\int \frac{\mathrm{d}%
q^{0}}{2\pi }\prod\limits_{a=1}^{{N}}\frac{2\omega _{a}}{(q^{0}+k^0_{a})^{2}-%
\omega _{a}^{2}+i\epsilon_a},  \qquad \omega_a=\sqrt{(\boldsymbol{q}-\boldsymbol{k}_a)^2+m^2_a},
\end{equation}
where $k_a$ are the external momenta\footnote{In order to have more symmetric formulas we have introduced one momentum for each external leg, which is redundant.}. With this definition, a standard Feynman diagram is
\begin{equation}
G_{N}=\int \frac{\mathrm{d}^{D-1}\boldsymbol{q}}{(2\pi )^{D-1}}\left(
\prod\limits_{a=1}^{N}\frac{1}{2\omega _{a}}\right) G_{N}^{s}.
\end{equation}
The skeleton diagrams satisfy the  identity~\cite{Anselmi:2021hab}
\begin{equation}\label{eq:spectralopt}
G^s+(G^s)^*=-\sum_{\text{cuts}}G^s_c,
\end{equation}
where $G^s_c$ are the cut skeleton diagrams. Moreover, each diagram in~\eqref{eq:spectralopt} can be decomposed into a sum of independent terms, which means that~\eqref{eq:spectralopt} gives a set of identities that holds independently. The decomposition works as follows. First, perform the integral over the loop energies by means of the residue theorem. After this operation, the result has the form
\begin{equation}
G^s_N=\sum_ic'_i\prod_{j=1}^N\frac{1}{D'_{ij}},
\end{equation}
where the coefficients $c_i'$ can depend on the spatial external momenta and the denominators $D'$ are linear combinations of external energies $k_a^0$ and frequencies $\omega'_a=\omega_a-i\epsilon_a$. It is always possible to manipulate the above expression such that the denominators contains only sums of frequencies, i.e.
\begin{equation}\label{eq:skeletonred}
    G_N^s=\sum_ic_i\prod_{j=1}^N\frac{1}{D_{ij}}, \qquad D_{ij}=P^0-\sum_a \omega'_a,
\end{equation}
where $P^0$ is a combination of external energies $k^0_a$. This means that the true physical nonanalyticities are always associated to thresholds which Lorentz invariant expressions have the form
\begin{equation}
P^2\geq \left(\sum_im_i\right)^2,
\end{equation}
where $P^2$ is some invariant build with external momenta and $m_i$ are internal masses. Thresholds that contains differences of masses are called \emph{pseudothresholds} and they need to disappear from the amplitudes in order to have a physically consistent theory. In fact, pseudethresholds are responsible for instabilities. For example, if the pseudothreshold
\begin{equation}
    P^2\geq (m_1-m_2)^2,
\end{equation}
is present, then a particle with squared mass $P^2=m^2$ would be allowed to decay into two heavier particles with masses $m_1, m_2 > m$ as long as their difference satisfies the bound.

The absence of pseudothresholds is nontrivial and depends on the choice of the prescription. In fact, the location of the poles in the loop-energy complex plane might change with different prescriptions and the integration contour used in the residue theorem would enclose different poles. For example, choosing the Feynman prescription for each degrees of freedom fixes the relative sign between $\omega_a$ and $i\epsilon_a$ to be the same in every $\omega'_a$. In general, mixing the prescription could change that. An example is given by mixing Feynman and anti-Feynman prescriptions in the same diagram by choosing the former for some degree of freedom and the latter for others. In that case, some of the physical thresholds are switched with pseudothresholds~\cite{Anselmi:2020tqo}. From what follows, it is clear that mixing the Feynman prescription and the fakeon prescription does not lead to instabilities and the pseudothresholds still cancel.

Once a skeleton diagram is reduced in the form~\eqref{eq:skeletonred} we apply the formula
\begin{equation}\label{eq:feynlim}
\lim_{\epsilon\rightarrow 0^+}\frac{1}{x+i\epsilon}=\mathcal{P}\frac{1}{x}-i\pi\delta (x),
\end{equation}
to each term, where the limit is understood in the sense of distributions and $\mathcal{P}$ is the Cauchy principal value. At this level each threshold, or at least what would be a threshold once we finalize the integral over the space components of loop momenta, is associated with a delta function. It is important to separate the terms that contain independent thresholds. For this purpose we introduce a few definitions in order to rewrite the skeleton diagram. We define
\begin{equation}
\mathcal{P}^{ab}=\mathcal{P}\frac{1}{e_{a}-e_{b}-\omega
_{a}-\omega_{b}}, \quad \mathcal{Q}^{ab}=\mathcal{P}\frac{2\omega _{b}}{(e_{a}-e_{b}-\omega
_{a})^2-\omega^2_{b}},\quad
\Delta ^{ab}=\pi \delta (e_{a}-e_{b}-\omega _{a}-\omega_{b}),
\label{eq:PQDdef}
\end{equation}
where $e_a\equiv k^0_a$ and the subscripts $a$, $b$ etc... label the internal legs. For example the skeleton bubble and skeleton triangle diagrams are
\begin{equation}
    B^s=-i\mathcal{P}_2-\Delta ^{12}-\Delta ^{21}
\end{equation}
\begin{equation}
C^{s}=-i\mathcal{P}_{\text{3}}+\sum_{\text{perms}}\left[-\Delta ^{ab}%
\mathcal{Q}^{ac}+\frac{i}{2}\Delta ^{ab}(\Delta
^{ac}+\Delta ^{cb})\right],  \label{Tdecomp}
\end{equation}%
respectively, where
\begin{equation}
    \mathcal{P}_2=\mathcal{P}^{12}+\mathcal{P}^{21}, \qquad \mathcal{P}_3=\mathcal{P}^{12}\mathcal{P}^{13}+\text{cycl}+(e\rightarrow -e).
\end{equation}
In general the quantity $\mathcal{P}_n$ is a sum of products of $n-1$ different $\mathcal{P}^{ab}$ and always contains the ultraviolet divergences of the associated diagram.
\begin{table}[t]
\begin{center}
\newcolumntype{P}[1]{>{\centering\arraybackslash}p{#1}}
\newcolumntype{M}[1]{>{\centering\arraybackslash}m{#1}}
 \begin{tabular}{|P{1.3cm}|M{1.2cm}|M{1.2cm}|M{1.2cm}|M{1.2cm}|}
 \hline
 \diagbox[innerwidth=1.3cm,height=1.5cm]{\small Terms}{\small Diag.} & \AB & \ApBp  & \ABp & \ApB\\ \hline 
$\mathcal{P}_{2}$ & $-i$ & $i$ & $0$ & $0$ \\ 
\hline
$\Delta ^{12}$ & $-1$ & $-1$ & $2$ & $0$ \\ \hline
$\Delta ^{21}$ & $-1$ & $-1$ & $0$ & $2$ \\ \hline
\end{tabular}
\end{center}
\caption{Threshold decomposition of the bubble diagram and its cut diagrams.}
\label{tab:tb1}
\end{table}%

Each Feynman diagram, as well as the correspondent cut diagrams, can be decomposed in this way (see~\cite{Anselmi:2021hab} for a general strategy). It is useful to show these results in a table. The case of the bubble diagram is depicted in~\autoref{tab:tb1}, where each column below the diagrams shows the coefficients of the decomposition that multiply the terms in the first column. The sum of the diagrams in the table vanishes due to the cutting identities. However, it is clear from the table that each row cancel independently. This new set of identities are called \emph{spectral identities}. In general, the independent terms that compose a skeleton diagram can be separated by the number of $\Delta$ that they contain. Indeed, in more complicated diagrams different products of $\Delta$'s appear. The threshold decomposition of a $L$-loop diagram with $N$ legs $G_N^{(L)}$ and its cut diagrams $G_N^{(L,n)}$ can be schematically written as
\begin{equation}
    G_N^{(L)}=-i\mathcal{P}_N^{(L)}+\sum_{j}\sum_mc_j\mathcal{O}_{j\Delta,m}
\end{equation}
\begin{equation}
    G_N^{(L,n)}=\sum_{j}\sum_mc_j^{(n)}\mathcal{O}_{j\Delta,m},
\end{equation}
where $\mathcal{O}_{j\Delta,m}$ are terms that contain $j$ number of $\Delta$'s, $m$ labels the number of terms with the same $j$, $n$ labels the number of cut diagrams and $c_j$, $c_j^{(n)}$ are constant coefficients. The spectral identities then read
\begin{equation}\label{eq:spectid}
    c_j+c_j^*+\sum_n c_j^{(n)}=0, \qquad \forall j.
\end{equation}
Moreover, a few other properties can be derived. First, by construction, the coefficients with even number of $\Delta$'s are purely imaginary, while those with an odd number are real, i.e.
\begin{equation}
    \text{Re}\left[c_{2r}\right]=\text{Re}\left[c_{2r}^{(m)}\right]=0, \qquad \text{Im}\left[c_{2r+1}\right]=\text{Im}\left[c_{2r+1}^{(m)}\right]=0, \qquad \forall r,m.
\end{equation}
Then, from this property and the spectral identities it follows that 
\begin{equation}
    \sum_{m}c_{2r}=0, \qquad \forall r.
\end{equation}
\begin{table}[t]
\begin{center}
\setlength{\tabcolsep}{3pt}
\renewcommand{\arraystretch}{1.1}
\newcolumntype{M}[1]{>{\centering\arraybackslash}m{#1}}
\newcolumntype{P}[1]{>{\centering\arraybackslash}p{#1}}
\begin{tabular}{|P{1.5cm}|M{1.5cm}|M{1.3cm}|M{1.3cm}|M{1.3cm}|M{1.3cm}|M{1.3cm}|M{1.3cm}|M{1.3cm}|M{1.3cm}|M{1.3cm}|}
\hline
\diagbox[innerwidth=1.5cm,height=1.5cm]{\small Terms}{\small Diag.}  & \ABC & \ApBpCp & \ABpCp & \ApBCp & \ApBpC
& \ApBC & \ABpC & \ABCp \\ \hline
$\mathcal{P}_{3}$ & $-i$ & $i$ & $0$ & $0$ & $0$ & $0$ & $0$ & $0$ \\ \hline
$\Delta^{12}\mathcal{Q}^{13}$ & $-1$ & $-1$ & $2$ & $0$ & $0$ & $0$ & $0$ & $0$ \\ \hline
$\Delta^{23}\mathcal{Q}^{21}$ & $-1$ & $-1$ & $0$ & $2$ & $0$ & $0$ & $0$ & $0$ \\ \hline
$\Delta^{31}\mathcal{Q}^{32}$ & $-1$ & $-1$ & $0$ & $0$ & $2$ & $0$ & $0$ & $0$ \\ \hline
$\Delta^{21}\mathcal{Q}^{23}$ & $-1$ & $-1$ & $0$ & $0$ & $0$ & $2$ & $0$ & $0$ \\ \hline
$\Delta^{32}\mathcal{Q}^{31}$ & $-1$ & $-1$ & $0$ & $0$ & $0$ & $0$ & $2$ & $0$ \\ \hline
$\Delta^{13}\mathcal{Q}^{12}$ & $-1$ & $-1$ & $0$ & $0$ & $0$ & $0$ & $0$ & $2$ \\ \hline
$\Delta^{12}\Delta^{13}$ & $i$ & $-i$ & $2i$ & $0$ & $0$ & $0$ & $0$ & $-2i$ \\ \hline
$\Delta^{23}\Delta^{21}$ & $i$ & $-i$ & $0$ & $2i$ & $0$ & $-2i$ & $0$ & $ 0 $ \\ \hline
$\Delta^{31}\Delta^{32}$ & $i$ & $-i$ & $0$ & $0$ & $2i$ & $0$ & $-2i$ & $ 0 $ \\ \hline
$\Delta^{21}\Delta^{31}$ & $i$ & $-i$ & $0$ & $0$ & $2i$ & $-2i$ & $0$ & $ 0 $ \\ \hline
$\Delta^{32}\Delta^{12}$ & $i$ & $-i$ & $2i$ & $0$ & $0$ & $0$ & $-2i$ & $0$ \\ \hline
$\Delta^{13}\Delta^{23}$ & $i$ & $-i$ & $0$ & $2i$ & $0$ & $0$ & $0$ & $-2i $ \\ \hline
\end{tabular}%
\end{center}
\caption{Threshold decomposition of the triangle diagram and its cut diagrams.}
\label{tab:ts1}
\end{table} 
These properties can be better appreciated in the case of the triangle diagram shown in~\autoref{tab:ts1}. 

The spectral identities~\eqref{eq:spectid} and the threshold decomposition allow us to reduce the optical theorem to a set of algebraic equations. To summarize, the sum of the spectral identities of a given diagram gives the cutting identities for that diagram, while the sum of the cutting identities implies the optical theorem. As mentioned in~\autoref{sec:unit}, the last implication is true only if some conditions are satisfied, for example, if all the degrees of freedom have positive residue. In the case of gauge theories it is necessary to include Faddeev-Popov ghosts in order to obtain unitarity in the Fock subspace where they are projected away, together with the longitudinal and temporal components of the gauge fields. On the other hand, theories with ghosts satisfy the cutting identities but violate the optical theorem. However, thanks to the threshold decomposition and the spectral identities we know how to modify the amplitudes and obtain a unitary theory in the subspace where some degrees of freedom are removed from the states that can be on shell. This is achieved by simply setting to zero all the $\Delta$'s that contain at least one frequency associated to the particles that  we want to remove from the spectrum, i.e. $\Delta^{ab}=0$ if the leg $a$ and/or $b$ is purely virtual. Indeed it is easy to see from~\autoref{tab:tb1} and~\autoref{tab:ts1} that if we cancel any row we would just remove one of the spectral identities, which hold independently. 
\begin{table}
\begin{center}
\newcolumntype{M}[1]{>{\centering\arraybackslash}m{#1}}
\newcolumntype{P}[1]{>{\centering\arraybackslash}p{#1}}
 \begin{tabular}{|P{1.3cm}|M{1.2cm}|M{1.2cm}|M{1.2cm}|M{1.2cm}|M{1.2cm}|}
 \hline
 \diagbox[innerwidth=1.3cm,height=1.5cm]{\small Terms}{\small \,Diag.} & \ABC & \ApBpCp & \ApBCp & \ABpC \\ \hline
 $\mathcal{P}_{3}$ & $-i$ & $i$ & $0$ & $0$\\ \hline
 $\Delta^{23}\mathcal{Q}^{21}$ & $-1$ & $-1$ & $2$ & $0$ \\ \hline
 $\Delta^{32}\mathcal{Q}^{31}$ & $-1$ & $-1$ & $0$ & $2$ \\ \hline
 \end{tabular}
 \end{center}
 \caption{The threshold decomposition of a triangle diagram and its cut diagrams where particle 1 is purely virtual.}
\label{tab:trifake}
 \end{table}
For example, \autoref{tab:ts1} reduces to~\autoref{tab:trifake} if particle 1 is purely virtual. Note that, after setting to zero every $\Delta^{1,a}$ also some of cut diagrams vanishes (those where particle-1 leg is cut). In the end the surviving rows still sum to zero. Therefore, when ghosts are present, if we remove the rows that are responsible for the violation of the optical theorem, we get a modified set of cutting identities which imply unitarity, once we project onto the subspace where the ghosts are not external states. In other words, the fakeon prescription allows us to consistently remove the ghosts (or any degrees of freedom we decide to) from the physical spectrum without relying on a symmetry and without changing the properties under renormalization (we never cancel the row that contains $\mathcal{P}_n$). Note that this removal is a true one and it is understood at any energy. This makes purely virtual particles radically different from resonances or unstable particles, for which always exists a Lorentz frame where they can be long lived and therefore detectable as every other particle.

In practice, the operation just described can be implemented by simply changing~\eqref{eq:feynlim} into
\begin{equation}\label{eq:fakelim}
\frac{1}{x+i\epsilon}\rightarrow\mathcal{P}\frac{1}{x}-i\tau\pi\delta (x),
\end{equation}
where $\tau=0$ if $x$ contains at least one fakeon frequency and $\tau=1$ otherwise. Then we can safely consider only diagrams where the purely virtual particles do not appear in the external legs. This is the fakeon prescription, which can be applied to both ghosts and standard particles and is consistent at any loop order~\cite{Anselmi:2018kgz,Anselmi:2021hab}. 

A given amplitude $\mathcal{A}_{\text{f}}$ in theories with fakeons, which is obtained using~\eqref{eq:fakelim}, can be written as 
\begin{equation}\label{eq:ampmod}
    \mathcal{A}_{\text{f}}=\mathcal{A}-\Delta_{\text{f}}\mathcal{A},
\end{equation}
where $\mathcal{A}$ is a standard amplitude obtained using the Feynman prescription and $\Delta_{\text{f}}\mathcal{A}$ is a functions obtained by summing all the terms that need to be set to zero by the fakeon prescription. This subtraction has been explicitly calculated in~\cite{Melis:2022tqz} for the case of one-loop bubble, triangle and box diagrams, which are the most relevant in particle physics phenomenology. The modified functions exhibit new nonanalyticities and singularities that can be used to discriminate models with fakeons from models without. It is even possible to explore the possibility that fakeons might exist in general, regardless the problem of ghosts in quantum gravity. For example, in~\cite{Anselmi:2021icc} a inert-doublet model where the new scalars are turned into fakeons has been considered as an extension of the standard model and compared with the standard inert-doublet model. The results show that in some portion of the parameters space the two models can be quite different. In particular, it was shown that the contribution to the decay of the Higgs boson into two photons differs in the two cases due to the modifications of the form~\eqref{eq:ampmod}. Another possibility is to consider whether one of the particles in the standard model can be purely virtual or not. The only particle that cannot be ruled out from being a fakeon by the present experimental data is the Higgs boson~\cite{Anselmi:2018yct}. This possibility is more interesting because there is no freedom in the parameters space and therefore we can look for some energy domains where the differences between the two cases (standard Higgs versus purely virtual Higgs) are relevant. This will be published in a forthcoming paper~\cite{PivaMelis}. 

The fakeon prescription opens the way for a new understanding of model building and particle physics. However, its main application remains quantum gravity. Although in principle quantum gravity with purely virtual particles can be discriminated from Stelle gravity or Einstein theory in terms of scattering amplitudes, it is quite challenging from the practical point of view, given the energies in play. Instead, a good arena where to test quantum gravity is inflationary cosmology. In the next section we show how to obtain an important prediction from quantum gravity in that context.

\sect{Quantum gravity with purely virtual particles and cosmology}
\label{sec:QGcosm}

In this section we write the action~\eqref{lhd} in an alternative form, in order to make clear the presence of the ghost and show how to apply the fakeon prescription. Then we proceed with the derivation of observables in inflationary cosmology and derive a prediction for the tensor-to-scalar ratio.

The procedure to rewrite the action is a composition of a Weyl transformation and a metric field redefinition, together with the introduction of scalar and spin-2 auxiliary fields $\phi$ and $\chi_{\mu\nu}$, respectively. The details can be found in~\cite{Anselmi:2018tmf}. The result of this procedure is that the action~\eqref{lhd} can be rewritten as
\begin{equation}
S_{\text{HD}}(g,\phi ,\chi)=\tilde{S}_{\text{HE}}(g)+S_{\chi }(g,\chi
)+S_{\phi }(g+\psi ,\phi ),  \label{sew}
\end{equation}%
where
\begin{equation}
    \tilde{S}_{\text{HE}}(g) =-\frac{1}{2\kappa ^{2}}\int \sqrt{-g}\left( 2%
\tilde{\Lambda}_{C}+\tilde{\zeta}R\right), \qquad \frac{\tilde{\Lambda}_{C}}{\Lambda _{C}}=\left( 1+\frac{2}{3}%
\frac{(\alpha +2\xi )\Lambda _{C}}{\zeta ^{2}}\right) ,\qquad \tilde{\zeta}=\zeta \frac{\tilde{%
\Lambda}_{C}}{\Lambda _{C}},
\end{equation}
\begin{equation}
S_{\phi }(g,\phi )=\frac{3\hat{\zeta}}{4}\int \sqrt{-g}\left[ \nabla_{\mu }\phi
\nabla^{\mu }\phi -\frac{m_{\phi }^{2}}{\kappa ^{2}}\left( 1-\mathrm{e}^{\kappa
\phi }\right) ^{2}\right] ,\qquad \hat{\zeta}=\zeta \left( 1+\frac{4}{3}\frac{\xi \Lambda _{C}}{%
\zeta ^{2}}\right) ,\qquad  \label{sphi}
\end{equation}
and $S_{\chi}(g,\chi)$ is the covariantized Pauli-Fierz action with the ``wrong" sign for the kinetic term plus nonminimal couplings and infinite interactions between $g$ and $\chi$, i.e.
\begin{equation}
S_{\chi }(g,\chi )=-\frac{\tilde{\zeta}}{\kappa ^{2}}S_{\text{PF}}(g,\chi
,m_{\chi }^{2})-\frac{\tilde{\zeta}}{2\kappa ^{2}}\int \sqrt{-g}R^{\mu \nu
}(\chi_{ \ \sigma}^{\sigma} \chi _{\mu \nu }-2\chi _{\mu \rho }\chi _{ \ \nu }^{\rho })+S_{\chi
}^{(>2)}(g,\chi ),  \label{scc}
\end{equation}
\begin{eqnarray}
S_{\text{PF}}(g,\chi ,m_{\chi }^{2}) &=&\frac{1}{2}\int \sqrt{-g}\left[
\nabla_{\rho }\chi _{\mu \nu }\nabla^{\rho }\chi ^{\mu \nu }-\nabla_{\rho }\chi_{ \ \mu}^{\mu} \nabla^{\rho
}\chi_{ \ \nu}^{\nu} +2\nabla_{\mu }\chi^{\mu \nu }\nabla_{\nu }\chi_{ \ \rho}^{\rho} -2\nabla_{\mu }\chi ^{\rho \nu
}\nabla_{\rho }\chi _{\nu }^{\mu }\right.  \notag \\
&&\qquad \left. -m_{\chi }^{2}(\chi _{\mu \nu }\chi ^{\mu \nu }-\chi_{ \ \mu}^{\mu}\chi_{ \ \nu}^{\nu})\right]
\label{SPF}
\end{eqnarray}%
where $S_{\chi }^{(>2)}(g,\chi )$ contains
terms that are at least cubic in $\chi$. They can be derived from the Einstein-Hilbert action as
\begin{equation}
S_{\chi }(g,\chi )=\tilde{S}_{\text{HE}}(g+\psi )-\tilde{S}_{\text{HE}%
}(g)-2\int \chi _{\mu \nu }\frac{\delta \tilde{S}_{\text{HE}}(g+\psi )}{%
\delta g_{\mu \nu }}+\frac{\tilde{\zeta}^{2}}{2\alpha \kappa ^{2}}\int
\left. \sqrt{-g}(\chi _{\mu \nu }\chi ^{\mu \nu }-\chi_{ \ \mu}^{\mu}\chi_{ \ \nu}^{\nu})\right\vert
_{g\rightarrow g+\psi },  \label{spsi}
\end{equation}%
and 
\begin{equation}
    \psi_{\mu\nu}=2\chi _{\mu \nu }+\chi _{\mu \nu }\chi_{ \ \rho}^{\rho}
-2\chi _{\mu \rho }\chi _{\nu }^{\rho }.
\end{equation}

Finally, the squared masses of $\phi$ and $\chi$ read
\begin{equation}
    m_{\phi}^2=\frac{\zeta}{\xi}, \qquad m_{\chi}^2=\frac{\tilde{\zeta}}{\alpha}.
\end{equation}
We highlight that the tilde and hat quantities are useful to deal with a nonvanishing cosmological constant and derive the correct change of variables that makes only $m_{\chi}$ modified by $\Lambda_C$. In the case where the cosmological constant is negligible, all the tilde and hat quantities are equal to the usual ones and it is convenient to choose
\begin{equation}
\psi_{\mu\nu}=2\chi_{\mu\nu}.
\end{equation}
For details, see~\cite{Anselmi:2018tmf}.

Now that all the degrees of freedom are manifestly represented by fields, it is clear how to apply the fakeon prescription: in every diagram we set to zero all the $\Delta$'s that contain a frequency associated to $\chi$. Then we restrict only to diagrams with external $h_{\mu\nu}$ and/or $\phi$ legs. In general we can couple any type of matter $\Phi$ to quantum gravity and the interactions in the variables~\eqref{sew} are obtained by means of the substitution
\begin{equation}
   S_{m}(g,\Phi)\rightarrow S_{m}(g\mathrm{e}^{\kappa \phi }+\psi 
\mathrm{e}^{\kappa \phi },\Phi ),
\end{equation}
where $S_m$ is the action of matter.

We stress that the action~\eqref{lhd} [or~\eqref{sew}] is an interim one. This means that it is not the action that we would obtain in the classical limit. Roughly speaking, the reason is that the fakeon is a purely quantum object and does not have a classical counterpart. Therefore, the fakeon prescription and all its physical effects cannot be derived from a classical action. We first need to start from a quantum theory with the desired properties and then perform the classical limit. We call the true classical action the \emph{projected action} and the interim one the \emph{unprojected action}. Then, one could wonder why we dot not start directly from the projected action, where the fields $\chi$ are already projected away, and quantize it with standard techniques. There are various reasons for this. First, it is very hard to obtain it explicitly, since the fakeon prescription is derived from perturbative quantum field theory, the whole procedure is also perturbative and it is not known at the moment how to implement it at a nonperturbative level. Moreover, the projected nonperturbative action is expected to be nonlocal and rather cumbersome. The good of using action~\eqref{lhd} is that it is local and we can define Feynman rules as usual (although we eventually modified the results as explained in~\autoref{sec:unit}). We could think of this in a reversed way as follows. Assume that a nonperturbative projected action of~\eqref{sew} exists and it is known explicitly. This action, $S_{\text{np}}^{\text{proj}}(g,\phi)$,  would be dependent only on the metric and the scalar field $\phi$ (and on other physical fields, if present). Moreover, it would be of a form where renormalizability is not manifest and it is complicated (or even impossible) to define Feynman rules. The fakeon prescription tells us that it is possible to introduce a new field $\chi$ and define a new action $S(g,\phi,\chi)$, which is physically equivalent to $S_{\text{np}}^{\text{proj}}$, provided that the fakeon prescription is used for $\chi$. The new action is~\eqref{sew} and such that
\begin{equation}
    S(g,\phi,\chi(g,\phi))=S_{\text{np}}^{\text{proj}}(g,\phi),
\end{equation}
where $\chi(g,\phi)$ is obtained from the nonperturbative version of the fakeon prescription. The advantage of $S(g,\phi,\chi)$ is that it is local. Moreover, being also equivalent to~\eqref{lhd}, it can be proved to be renormalizable, as explained in~\autoref{sec:Stelle}. At the present it is possible to obtain only the perturbative version of projected action by studying the classical limit of the fakeon prescription, which is enough to work out predictions. Some explicit examples for simpler models have been obtained in~\cite{Anselmi:2019rxg}. The projected action in the case of gravity has been derived at quadratic level in the perturbations around inflationary background at several orders in the slow-roll expansion~\cite{Anselmi:2020lpp,Anselmi:2020neq} as we review below.

\subsection{Inflation}
The theory of quantum gravity with purely virtual particles can be tested in the context of inflationary cosmology. In fact, the scalar degree of freedom introduced by the $R^2$ term can be viewed as the inflaton and used to explain the anisotropies of the cosmic microwave background~\cite{Starobinsky:1980te}. Since our action contains also the $C^2$ term, we need to treat it in a fashion similar to what is explained in~\autoref{sec:unit} in the case of scattering amplitudes. For this reason, it is necessary to understand the fakeon prescription in curved spacetime. As explained below, this task is simplified by the fact that in cosmology we do not need to go as far as computing loop corrections. Therefore, we can work with the classical limit of the fakeon
prescription~\cite{Anselmi:2020lpp}.

It is more convenient to use the action
\begin{equation}\label{eq:Sinf}
    S(g,\phi)=-\frac{M_{\text{Pl}}^2}{16\pi^2}\int\sqrt{-g%
}\left(R+\frac{1}{2m_{\chi }^{2}}C_{\mu \nu \rho \sigma }C^{\mu \nu \rho
\sigma }\right)+\frac{1}{2}\int\sqrt{-g}\left[\nabla_{\mu
}\phi \nabla^{\mu }\phi -2V(\phi )\right], 
\end{equation}
\begin{equation}
V(\phi )=\frac{m_{\phi }^{2}}{2\hat{\kappa}^{2}}\left( 1-\mathrm{e}^{\hat{%
\kappa}\phi }\right) ^{2}, \qquad \hat{\kappa}=M_{\text{Pl}}^{-1}\sqrt{16\pi /3},  \label{staropot}
\end{equation}%
where we have explicitly introduced the Planck mass $M_{\text{Pl}}^2=8\pi^2\zeta/\kappa^2$, the fakeon mass $m_{\chi}^2=\zeta/\alpha$ and the inflaton mass $m_{\phi}^2=\zeta/\xi$, and we have canonically normalized the field $\phi$. Moreover, we set the cosmological constant to zero since it is unimportant for our purposes. Alternatively, it is possible to use directly the action~\eqref{lhd}. Here we consider only the action~\eqref{eq:Sinf}. The two possibilities are physically equivalent and they are connected by perturbative redefinitions of the quantities~\cite{Anselmi:2020lpp}.

Before to proceed we fix some notation. We expand the action~\eqref{eq:Sinf} around the Friedmann-Lema\^itre-Robertson-Walker (FLRW) background
\begin{equation}
g_{\mu\nu}=\bar{g}_{\mu\nu}+\delta g_{\mu\nu} \qquad \bar{g}_{\mu \nu }=\text{diag}(1,-a^{2},-a^{2},-a^{2})
\end{equation}
up to the quadratic order in the perturbations $\delta g_{\mu\nu}$, where $a(t)$ is the scale factor. We anticipate to the reader that in the end only the tensor and scalar perturbations propagate and no vector or additional tensors/scalars are present due to the fakeon projection. Therefore, instead of going through the details for the two cases separately, we show the general procedure which is valid for both (specifying the differences where necessary) and then present the power spectra and the spectral indices.

We work in space Fourier transform and label the space momentum as $\mathbf{k}$ and its modulus as $|\mathbf{k}|=k$. We define the slow-roll parameter 
\begin{equation}
    \varepsilon\equiv-\frac{\dot{H}}{H^2}, \qquad H=\frac{\dot{a}}{a}
\end{equation}
and express every quantities as power series in $\varepsilon$, up to overall, non polynomials factors. In particular, it is easy to show that 
\begin{equation}
\frac{\mathrm{d}^{n}\varepsilon }{\mathrm{d}t^{n}}=H^{n}\mathcal{O}%
(\varepsilon ^{\frac{n+2}{2}})  \label{assum}
\end{equation}%
so other slow-roll parameters that are typically defined in the literature, such as $\eta\equiv 2\varepsilon -\frac{\dot{%
\varepsilon}}{2H\varepsilon}$, are also series in $\varepsilon$. In general, inflationary models contain one independent slow-roll parameter for each field that participates to inflation. It is also useful to derive the expansions 
\begin{eqnarray}\label{eq:Hv}
H &=&\frac{m_{\phi }}{2}\left( 1-\frac{\sqrt{3\varepsilon }}{2}+\frac{%
7\varepsilon }{12}+\mathcal{O}(\varepsilon ^{3/2})\right) ,  \notag \\
v&\equiv&-aH\tau =1+\varepsilon +\mathcal{O}(\varepsilon ^{3/2}).  \notag
\end{eqnarray}%
where $\tau $ is the conformal time
\begin{equation}
\tau =-\int_{t}^{+\infty }\frac{\mathrm{d}t^{\prime }}{a(t^{\prime })},
\label{tau}
\end{equation}%
with the initial condition chosen to have $\tau =-1/(aH)$ in the de Sitter limit $\varepsilon \rightarrow 0$. Moreover, it is convenient to define
\begin{equation}\label{eq:lambda}
    \lambda\equiv\frac{\hat{\kappa}\dot{\phi}}{2 H}=\sqrt{-\frac{\dot{H}}{3H^2}}=\sqrt{\frac{\varepsilon}{3}},
\end{equation}
since every quantity is an expansion in $\sqrt{\varepsilon}$ (and eventually also in $k$). The parameter $\lambda$ is small during inflation and can be viewed as a ``coupling constants" from which the whole process of cosmic inflation is interpreted as a sort of ``renormalization-group flow"~\cite{Anselmi:2020shx}, in analogy with particle physics. This is a mathematical correspondence between the quantities in inflation and in perturbative quantum field theory and it is useful to systematize the computations by exporting the techniques of the latter to the former. This idea is generalized for several single-field inflationary models~\cite{Anselmi:2021rye}, where potentials can be classified and even ruled out. The case of double field inflation is studied in~\cite{Anselmi:2021dag}. For future use we write the background field equations in terms of $\lambda$. The Friedmann equations are
\begin{equation}
\dot{H}=-\frac{3\hat{\kappa}^2}{4}\dot{\phi}^{2},\qquad 
H^2=\frac{\hat{\kappa}^2}{4}\left(\dot{\phi}^{2}+2V\right) ,\qquad \ddot{\phi}+3H\dot{\phi}+\frac{\mathrm{d}V}{\mathrm{d}\phi}=0.  \label{frie}
\end{equation}%
Then, by combining the equations~\eqref{frie}, it is easy to show that the parameter $\lambda$ satisfies
\begin{equation}\label{eq:lambdaeom}
    \frac{\dot{\lambda}}{H}=-3\lambda(1-\lambda^2)-\frac{\hat{\kappa}}{2H^2}\frac{\mathrm{d}V}{\mathrm{d}\phi}.
\end{equation}

Finally, we denote with $u_{\mathbf{k}}(t)$ the space Fourier transform of a general perturbation and the power spectrum of $\mathcal{P}_u$ is defined as
\begin{equation}
\langle u_{\mathbf{k}}(\tau )u_{\mathbf{k}^{\prime }}(\tau )\rangle =(2\pi
)^{3}\delta ^{(3)}(\mathbf{k}+\mathbf{k}^{\prime })\frac{2\pi ^{2}}{k^{3}}%
\mathcal{P}_{u}, \qquad \mathcal{P}_u=\frac{k^3}{2\pi^2}|u_{\mathbf{k}}|^2.  \label{deltau}
\end{equation}%
Moreover, replacing $|\tau |$ by $%
1/k_{\ast }$, where $k_{\ast }$ is a reference scale, we write 
\begin{equation}
\ln \mathcal{P}_{u}(k)=\ln A_{u}+n_{u}\ln \frac{k}{k_{\ast }},  \label{lnpr}
\end{equation}%
where $A_{u}$ and $n_{u}$ are called amplitude and spectral index, respectively. From now on we omit the subscript $\mathbf{k}$ in the perturbations and simply write $u(t)$. Terms like $u\dot{u}$ are understood as either $u_{\mathbf{k}}\dot{u}_{-\mathbf{k}}$ or $u_{-\mathbf{k}}\dot{u}_{\mathbf{k}}$.

In the case of tensor perturbations, the quadratic action is of the form
\begin{equation}
S(u)=\frac{M^2_{\text{Pl}}}{8\pi}\int \mathrm{d}t \ a(t)^3\left[f(t)\dot{u}^2-g(t)u^2-\frac{1}{m_{\chi}^2}\ddot{u}^2\right],
\end{equation}
where $f$ and $g$ are time-dependent functions. Note that, because of the $C^2$ term, there is a higher-derivative term, as expected. At this level, it is convenient to remove it with a field redefinition and introduce explicitly an additional perturbation. In order to do this we consider the extended action
\begin{equation}
S'(u,B)=S(u)+\Delta S, \quad \Delta S=\frac{M_{\text{Pl}}^2}{8\pi m_{\chi}^2}\int a^3\left(B-%
\ddot{u}-\tilde{f}\dot{u}-\tilde{g}u\right) ^{2},  \label{dlt}
\end{equation}%
where $B$ is an auxiliary field and $\tilde{f}$, $\tilde{g}$ are functions to be determined. The two actions coincide once we substitute $B$ with the solution of its own equation of motion $B(u)$, i.e.
\begin{equation}
    S'(u, B(u))=S(u).
\end{equation}
Finally, we perform the field redefinitions
\begin{equation}
u=U+b V,\qquad \qquad B=V+c U,  \label{uU}
\end{equation}
where $b$ and $c$ are other functions to be determined. In the case of scalars, there are no higher-derivative terms but some fields are not dynamical and can be removed algebraically by using their filed equations. 

After these procedures the action in both cases is reduced in the form  
\begin{equation}\label{eq:SUV}
 S^{\prime}(U,V)=\frac{1}{2}\int \mathrm{d}t \ Z\left(\dot{U}^2-\omega^2 U^2-\dot{V}^2+\Omega^2V^2+2\sigma UV\right),
\end{equation}
where $Z, \omega, \Omega$ and $\sigma$ are time-dependent functions that are different for tensor and scalar perturbations. It is now explicit that the field $V$ is problematic, due to the different sign of its kinetic term. Therefore, we need to quantize it as purely virtual and remove it from the spectrum. As mentioned above, we do not need to compute loops and the tree-level version of the fakeon prescription is enough. In order to obtain it we note that formula~\eqref{eq:fakelim} with $\tau=0$ can be written as
\begin{equation}\label{fakeonlim2}
    \frac{1}{2}\left(\frac{1}{x+i\epsilon}+\frac{1}{x-i\epsilon}\right)=\mathcal{P}\frac{1}{x}
\end{equation}
and apply it to the fakeon perturbations by averaging their retarded and advanced Green functions. The result of this operation is called \emph{fakeon Green function}. More explicitly, the equation of motion for $V$ is
\begin{equation}\label{eq:SigmaEom}
  \Sigma V\equiv  \left(\frac{\text{d}^2}{\text{d}t^2}+\frac{\dot{Z}}{Z}\frac{\text{d}}{\text{d} t}+\Omega^2\right)V=-\sigma U
\end{equation}
and the solution obtained with the fakeon Green function is
\begin{equation}
    V(t)=(G_{\text{f}}*F)(t), \qquad F\equiv -\sigma U,
\end{equation}
where $G_{\text{f}}$ is the fakeon Green function of the operator $\Sigma$ and ``$*$" is the convolution. The result for $G_{\text{f}}$ in de Sitter space is~\cite{Anselmi:2020lpp}
    \begin{equation}
G_{\text{f}}(t,t^{\prime })=\frac{i\pi \mathrm{sgn}(t-t^{\prime })\mathrm{e}%
^{-3H(t-t^{\prime })/2}}{4H\sinh \left( n_{\chi }\pi \right) }\left[
J_{in_{\chi }}(\check{k})J_{-in_{\chi }}(\check{k}^{\prime })-J_{in_{\chi }}(%
\check{k}^{\prime })J_{-in_{\chi }}(\check{k})\right],  \label{feGG}
\end{equation}
where $J_{n}$ denotes the Bessel function of the first kind and 
\begin{equation}
n_{\chi }=\sqrt{\frac{m_{\chi }^{2}}{H^{2}}-\frac{1}{4}},\qquad \check{k}=%
\frac{k}{a(t)H},\qquad \check{k}^{\prime }=\frac{k}{a(t^{\prime })H}.
\label{kappa}
\end{equation}
In this way we can write $V$ as a function of $U$
\begin{equation}\label{eq:V(U)}
    V(U)=-G_{\text{f}}*(\sigma U)
\end{equation}
and plug it back in the action~\eqref{eq:SUV}, obtaining the projected action
\begin{equation}\label{eq:projS}
    S^{\text{proj}}(U)=S'(U,V(U)).
\end{equation}
This is the classical version of fakeon prescription. It is worth note a few properties. First, the function $\sigma$ is of order $\lambda^2$~\cite{Anselmi:2020lpp} and, by~\eqref{eq:V(U)}, so is $V(U)$, and at the leading order in the slow-roll expansion the fakeon prescription gives $V=0$. Therefore, the nonlocal term $\sigma U V(U)$ in~\eqref{eq:projS} is $\mathcal{O}(\lambda^4)$, which means that the projected action is unaffected by $V(U)$ up to the order $\lambda^3$ included. This simplifies the computations, which can be pushed to higher-order without including the nonlocal contributions~\cite{Anselmi:2020neq}. However, the change of variables~\eqref{uU} tells us that we need $V(U)$ for the power spectra, since the true physical variable is $u(U,V)$. Moreover, the power spectra are computed in the so-called \emph{superhorizon limit} $|k\tau|\rightarrow 0$, so having $V(U)$ in that limit is enough. This further simplifies its derivation. The function $V(U)$ up to the order $\lambda^3$ is of the form
\begin{equation}
    V(U)=\lambda^2(v_1+v_2\lambda)U+v_3\lambda^3\dot{U}+\mathcal{O}(\lambda^4),
\end{equation}
where the coefficients $v_i$ depend on $m_\chi$ and $m_\phi$ and can be found in~\cite{Anselmi:2020neq}\footnote{Note that in~\cite{Anselmi:2020neq} the quantity~\eqref{eq:lambda} is labelled as $\alpha$, while here we have used the letter $\lambda$ to avoid confusion with the parameter in front of the $C^2$ term.}.

\subsection{Power spectra}\label{subsec:powerspectra}

At this stage we need only to compute the solution to the equation of motion for $U$ and then derive the power spectrum for $u$. In order to do it, we first change variable to a rescaled conformal time $\eta\equiv-\bar{k}\tau$, where $\bar{k}=k(1+\mathcal{O}(\lambda))$ is due to the the fakeon prescription, so $\bar{k}=k$ if the $C^2$ term is absent. Then we perform an additional field redefinition to put the projected action in the Mukhanov-Sasaki form
\begin{equation}\label{eq:MSact}
    S_w^{\text{proj}}(w)=\frac{1}{2}\int \mathrm{d}\eta \left[ \left(\frac{\mathrm{d}w}{\mathrm{d}\eta}\right)^2-w^{2}+\left(\frac{2+\sigma_{w}}{\eta^2
}\right)w^2 \right],
\end{equation}
where $w$ is the new variable, the $^\prime$ denotes the derivative with respect to $\eta$ and $\sigma_w=\mathcal{O}(\lambda)$ is a power series in $\lambda$ which encodes the deviations from scale-invariant power spectrum. The solution of the associated equation of motion can be derived by imposing the usual Bunch-Davies condition for the field $w$, which in these variables reads
\begin{equation}
    w(\eta)\simeq \frac{e^{i\eta}}{\sqrt{2}}, \qquad \text{for} \ \  \eta\rightarrow \infty.
\end{equation}
The solution is expanded in powers of $\lambda$ as
\begin{equation}
    w(\eta)=w_0(\eta)+\sum_{n=1}^{\infty}\lambda^nw_n(\eta),
\end{equation}
where 
\begin{equation}
    w_0(\eta)=\frac{i(1-i\eta)}{\eta\sqrt{2}}e^{i\eta}
\end{equation}
and $w_{n>0}$ are other functions that depend on the type of perturbations (tensor or scalar), since also $\sigma_u$ depends on that. In particular, 
$\sigma_u$ is $\mathcal{O}(\lambda)$ and $\mathcal{O}(\lambda^2)$, for scalar and tensor perturbations, respectively.

Finally, the power spectrum is obtained from $|u|^2$ by tracing back all the field redefinitions and change of variables starting from the solution $w(\eta)$, so we have
\begin{equation}
    u=U(w)+bV(U(w)), \qquad w=w(\bar{k}|\tau|)
\end{equation}
and the power spectrum is
\begin{equation}
    \mathcal{P}_u(k)=\frac{k^3}{2\pi^2}|u(k)|^2,
\end{equation}
It is important to highlight that in the superhorizon limit the dependence of $\mathcal{P}_u$ on $\tau$ drops and that the one on $k$ is all encoded in $\lambda_k\equiv\lambda(1/k)$, where $\lambda$ is obtained from the background equations of motion~\eqref{eq:lambdaeom} as a function of $\tau$ and then set $\tau=1/k$. The results for the power spectra for tensor and scalar perturbations are~\cite{Anselmi:2020lpp,Anselmi:2020neq}
\begin{equation}
    \mathcal{P}_t(k)=\frac{4m_{\phi}^2\varsigma}{\pi M_{\text{Pl}}^2}\left[1+3\varsigma\lambda_k+\lambda_k^2\left(6\varsigma\gamma_M+\frac{47}{4}\varsigma^2+\frac{11}{8}\frac{m_{\phi}^2}{m_{\chi}^2}\varsigma\right)+\mathcal{O}(\lambda_k^3)\right],\quad \varsigma\equiv\frac{1}{\left(1+\frac{m_{\phi}^2}{2m_{\chi}^2}\right)},
\end{equation}
\begin{equation}
    \mathcal{P}_s(k)=\frac{m_{\phi}^2}{12\pi M_{\text{Pl}}^2\lambda_k^2}\left[1+\lambda_k(5-4\gamma_M)+\lambda_k^2\left(4\gamma_M^2-\frac{40}{3}\gamma_M+\frac{7}{3}\pi^2-\frac{67}{12}-\frac{m_{\phi}^2}{2m_{\chi}^2}F\left(\frac{m_{\phi}^2}{m_{\chi}^2}\right)\right)+\mathcal{O}(\lambda_k^3)\right],
\end{equation}
respectively, where $\gamma_M=\gamma_E+\ln2$, $\gamma_E$ being the Euler-Mascheroni constant and $F$ is a function which can be obtained recursively as a series up to arbitrary orders~\cite{Anselmi:2020neq}.
The spectral indices are given by
\begin{equation}
    n_u-\theta=\frac{\mathrm{d}\ln\mathcal{P}_u(\lambda_k)}{\mathrm{d}\ln k}=-\beta(\lambda_k)\frac{\partial\ln \mathcal{P}_u}{\partial \lambda_k}, \qquad \beta(\lambda_k)=-\frac{\mathrm{d}\lambda_k}{\mathrm{d}\ln k},
\end{equation}
where $\theta=0,1$ for tensor and scalar perturbations, respectively. The \emph{beta function} $\beta$ is a power series in $\lambda_k$ and it is obtained from
\begin{equation}
\frac{\mathrm{d}\lambda}{\mathrm{d}\ln|\tau|}=-\frac{1}{v}\frac{\dot{\lambda}}{H}
\end{equation}
by using the expansions~\eqref{assum} and~\eqref{eq:Hv}. We report its expression up to the order $\lambda^4$
\begin{equation}\label{eq:betaexp}
\beta(\lambda)=-2\lambda^2\left[1+\frac{5}{6}\lambda+\frac{25}{9}\lambda^2+\mathcal{O}(\lambda^3)\right]
\end{equation}
but can be derived up to any order. In the renormalization-group-flow analogy mentioned above the quantity~\eqref{eq:betaexp} plays the role of the beta function in quantum field theory and a necessary condition for the background to be asymptotically de Sitter in the infinite past is that~\eqref{eq:betaexp} be at least quadratic and negative (in analogy with asymptotic freedom). 

Finally, the spectral indices read
\begin{equation}
    n_t=-6\varsigma\lambda_k^2+\mathcal{O}(\lambda_k^3), \qquad n_s-1=-4\lambda_k\left[1-\lambda_k\left(\frac{5}{3}-2\gamma_M\right)\right]+\mathcal{O}(\lambda_k^3).
\end{equation}
Note that in the case of scalar perturbations the contributions due to the fakeon (the $m_{\chi}$ dependence) start to appear at the order $\lambda_k^2$ for $\mathcal{P}_s$ and at the order $\lambda_k^3$ for $n_s$ (not shown here). This means that up to those orders the predictions of scalar perturbations are indistinguishable from those of the Starobinsky model. On the other hand, the predictions for the tensor perturbations get modified already at the leading order.

\subsection{Consistency condition}
The study of the fakeon prescription on curved spacetime leads to other nontrivial consequences. The whole procedure cannot be applied in the case of tachyons and specific conditions must be imposed. In flat spacetime, the so-called \emph{no-tachyon conditions} typically constrains the parameters in the action. For example, in the case of the quantum gravity action~\eqref{lhd} we must have $\alpha$, $\xi>0$ in order to avoid tachyons, i.e. $m_{\phi}^2$, $m_{\chi}^2>0$. In FLRW spacetime we need to impose that the mass squared of the field $V$ be positive, which can be read from~\eqref{eq:SigmaEom} after the redefinition that cancels the term with a single derivative, i.e.
\begin{equation}
    V\rightarrow \frac{V}{\sqrt{Z}}.
\end{equation}
Then the effective mass is 
\begin{equation}
    m(t)^2=\Omega^2+\frac{\dot{Z}^2}{4Z^2}-\frac{\ddot{Z}}{2Z},
\end{equation}
where the functions $\Omega$ and $Z$ are different for each perturbations. However, for tensor, vector and scalar perturbations it is of the form
\begin{equation}\label{eq:mt}
    m(t)^{2}=m_{\chi }^{2}-\frac{H^{2}}{4}+%
\frac{k^{2}}{a^{2}}+\mathcal{O}(\epsilon,k^4).
\end{equation}
It is enough to derive the no-tachyon condition in de Sitter spacetime, since we are expanding perturbatively around it. Moreover, we can further simplify the expression~\eqref{eq:mt} by taking the superhorizon limit $k/(aH)\rightarrow 0$. Then, recalling that $H(\varepsilon=0)=m_{\phi}/2$ and by imposing $m(t)^2\left.\right|_{k/(aH)\rightarrow 0}>0$ at $\varepsilon=0$ we get the consistency condition
\begin{equation}\label{eq:ABP}
m_{\chi}>\frac{m_{\phi}}{4}.
\end{equation}
It is possible to impose a stronger bound by requiring that $m(t)^2>0$ for every $k$. However, the positivity of a time-dependent function is not reparametrization invariant. This can be shown by considering the most general transformation $t\rightarrow t'(t)$,  $V(t)\rightarrow \hat{V}(t)$ that leaves the kinetic term invariant but changes the mass $m(t)^2$ into $M(t')^2$. Then, $m(t)^2>0$ does not necessarily imply $M(t')^2>0$. On the other hand, if $m^2$ is time independent and positive then the most general transformation that leaves $M^2$ $t'$-independent also leaves it positive~\cite{Anselmi:2020lpp}. Imposing the no-tachyon condition for every $k$ gives the same bound~\eqref{eq:ABP} in the case of tensor and vector perturbations and a stronger one in the case of scalar ones. However, in what follows we consider only the no-tachyon condition in the superhorizon limit and conclude that the presence of negative mass squared in some time interval is not necessarily a lack of consistency.

The bound~\eqref{eq:ABP} it is important for phenomenologial reasons, since, by combining it with experimental data, it narrows the allowed window for the tensor-to-scalar ratio to be less than an order of magnitude (see below).

\subsection{Predictions}
The best experimental data used to test inflationary models are those given by the Planck collaboration~\cite{Planck:2018jri} combined with those of the BICEP/Keck collaboation~\cite{BICEP:2021xfz}. In order to compare the data with the theoretical calculations we define the (dynamical) tensor-to-scalar ratio
\begin{equation}
    r(k)=\frac{\mathcal{P}_t(k)}{\mathcal{P}_s(k)}
\end{equation}
so that the usual tensor-to-scalar ratio is $r(k)$ at a reference scale $k_*$. From the results above we obtain
\begin{equation}
    r(k)=48\varsigma\lambda_k^2\left[1+\lambda_k(3\varsigma-5+4\gamma_M)\right]+\mathcal{O}(\lambda_k^3).
\end{equation}
The best data available at the moment give
\begin{equation}
    n_s(k_*)=0.9649\pm 0.0042,\qquad \ln\left(10^{10}\mathcal{P}_s(k_*)\right)=3.044\pm 0.014,\qquad  r(k_*)<0.035,
\end{equation}
where $k_*=0.05 \ \text{Mpc}^{-1}$. From the measurement of $n_s(k_*)$ we can extract the value of $\lambda_*\equiv \lambda_{k_*}$, while from $\mathcal{P}_s(k_*)$ we extract $m_{\phi}$. The results are
\begin{equation}
    \lambda_*=0.0087\pm 0.0010, \qquad m_{\phi}=(2.99\pm 0.36)\times10^{13} \ \text{GeV}.
\end{equation}
\begin{figure}
    \centering
    \includegraphics[width=17cm]{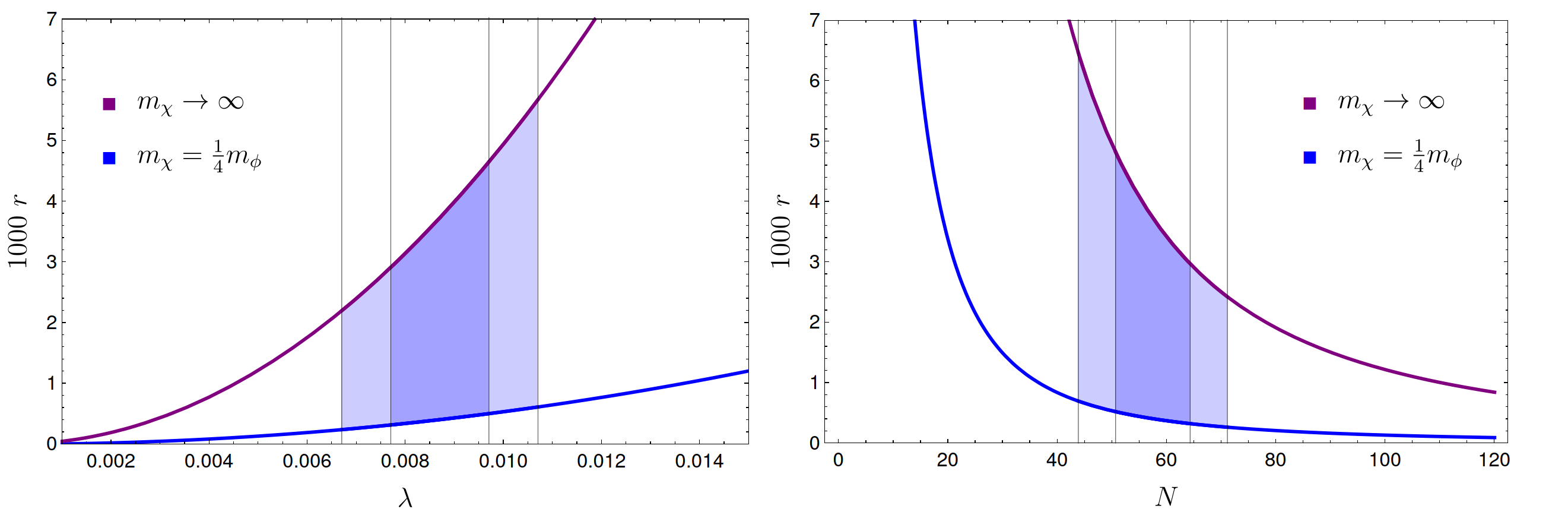}
    \caption{On the left panel, the tensor-to-scalar ratio as a function of the parameter $\lambda$. On the right panel the same quantity is plotted as a function of the number of e-folds. The purple lines represent $r$ in the Starobinsky model, while the blue lines represent $r$ when the consistency condition is saturated. The darker regions indicate the one-sigma level, while the ligher ones the two-sigma level.}
    \label{fig:rVSlambda}
\end{figure}
Then, knowing the allowed values for $\lambda_*$ and using the bound~\eqref{eq:ABP}, we can plot the tensor-to-scalar ratio as a function of $\lambda$ and restrict its values between the two curves in the cases $m_{\chi}=m_{\phi}/4$ and $m_{\chi}\rightarrow\infty$. The result up to two-sigma level is shown in the left panel of~\autoref{fig:rVSlambda}, while on the right panel we show it as a  function of the number of e-folds $N$
\begin{equation}
N=\int_{t_{i}}^{t_{f}}H(t^{\prime })\mathrm{d}t^{\prime },  \label{n}
\end{equation}%
which is often used in the literature to express the results. In general, it is more useful to use the variable $\lambda$, since the results can be expressed as power series of it (up to overall terms), while $N$ it is not a perturbative quantity. This follows from the relation between $N$ and $\lambda$
\begin{equation}
N=\int^{\frac{1}{\sqrt{3}}}_{\lambda}\frac{H(\lambda')}{\dot{\lambda}(\lambda')}\mathrm{d}\lambda'=\int^{\frac{1}{\sqrt{3}}}_{\lambda}\frac{\mathrm{d}\lambda'}{2\lambda'^2}\left[1-\frac{5}{6}\lambda'+\mathcal{O}(\lambda^{\prime 0})\right]=\frac{1}{2\lambda}+\frac{5}{12}\ln \lambda+\mathcal{O}(\lambda^0), \label{nlambda}
\end{equation}
From~\eqref{nlambda} we see that the $N(\lambda)$ is not a power series. The plots in~\autoref{fig:rVSlambda} show that in the theory of quantum gravity with purely virtual particles, where the field $\phi$ plays the role of the inflaton, the tensor-to-scalar ratio is confined in a window that is around an order of magnitude. For concreteness, if we take $N=60$ the tensor-to-scalar ratio is 
\begin{equation}
    0.37\lesssim1000 \ r\lesssim3.41.
\end{equation}
Future experiments, such as LiteBIRD~\cite{Hazumi:2019lys}, might be able to test this result, if the so-called $B$-modes are detected within the expected sensitivity ($\delta r<0.001$ for LiteBIRD). We highlight that by measuring one new quantity, such as $r(k_*)$ or $n_t(k_*)$, the parameter $m_{\chi}$ would be fixed and every other potential prediction would be a precision test of the theory.

\sect{Conclusions}
\label{sec:concl}
We have reviewed an approach to quantum gravity in the framework of quantum field theory. The theory is build by requiring the same guiding principles that have lead to the standard model, i.e. locality, renormalizability and unitarity, and by being as conservative as possible in accomodating those requirements. Renormalizability already singles out a unique action for quantum gravity, which, besides the massless graviton, contains a scalar that can be viewed as the inflaton, a cosmological constant and a massive spin-2 particle. The latter, if quantized by means of standard techniques, leads to a violation of unitarity. However, this degree of freedom is necessary to achieve rernomalizability and cannot be removed without loosing that property. In order to restore unitarity without renouncing to renormalizability we use a different quantization procedure for the massive spin-2 field, the fakeon prescription. Such procedure is very general and in principle can be applied to any degree of freedom. Effectively, the fakeon prescription amounts to compute every amplitude using the Feynman prescription as usual and then subtract certain functions, whose role is to remove the on-shell parts of the degrees of freedom that we want to quantize in this way. The outcome is that those particles become purely virtual and cannot appear on shell. In the case of quantum gravity, this is crucial since the possibility of having an on-shell spin-2 ghost violates unitarity. Using the fakeon prescription we can remove the ghost from the spectrum of the theory without loosing it from the possible virtual states. In this way we obtain a renormalizable and unitary theory of quantum gravity. We have shown that a good arena to test this theory is inflationary cosmology and shown that the consistency of the fakeon prescription in that context leads to a pretty sharp prediction for the tensor-to-scalar ratio, which could be tested in future experiments that measure the polarization of the cosmic microwave background.

When compared to other approaches to quantum gravity, the fakeon idea has several advantages. First it is perturbative and does not add more complications from the computational point of view. The effort in computing Feynman diagrams in the theory of quantum gravity with fakeons is comparable to that required for the standard model. Moreover, by reconciling renormalizability and unitarity in quantum gravity, the fakeon approach makes unnecessary to invoke nonperturbative approaches to renormalization, such as asymptotic safety. Furthermore, the whole procedure truly removes the ghost degrees of freedom from the theory, unlike the approach~\cite{Donoghue:2019fcb}, which proves unitarity in Stelle theory only in situations where the unstable ghost has decayed. In this regards, it is worth to highlight that ghosts can be removed by means of the fakeon prescription even if they are stable.

Finally, we mention some future directions in this research line. From the theoretical point of view it is important to understand some aspects of the theory of quantum gravity, for example its perturbative validity. In fact, the fakeon prescription applied to Stelle gravity introduces two power countings, so to speak. On the one hand the renormalization sector is perturbative up to arbitrary energies, since the power counting is that of Stelle gravity. On the other hand some quantities, such as absorptive parts of the amplitudes, obey the power counting of general relativity. An on-going work is to understand whether semi-nonperturbative techniques, combined with the fakeon prescription, can improve the nonrenormalizable behaviors of scattering amplitudes. On a more phenomenological side, it would be interesting to study the role of the spin-2 fakeon in post-inflationary eras to see how cosmology is impacted. Furthermore, the results of~\cite{Melis:2022tqz} need to be explored more deeply, to uncover other new effects in particle physics that could be directly related to the presence of fakeons.\\
\hfill\break
{\bf Data Availability Statement}: no data associated in the manuscript.

\bibliographystyle{JHEP} 
\bibliography{mybiblio}

\providecommand{\href}[2]{#2}\begingroup\raggedright\begin{thebibliography}{10}

\bibitem{Anselmi:2017yux}
D.~Anselmi and M.~Piva, {\it {A new formulation of Lee-Wick quantum field
  theory}},  {\em JHEP} {\bf 06} (2017) 066,
  [\href{http://arxiv.org/abs/1703.04584}{{\tt arXiv:1703.04584}}].

\bibitem{Anselmi:2017ygm}
D.~Anselmi, {\it {On the quantum field theory of the gravitational
  interactions}},  {\em JHEP} {\bf 06} (2017) 086,
  [\href{http://arxiv.org/abs/1704.07728}{{\tt arXiv:1704.07728}}].

\bibitem{tHooft:1974toh}
G.~'t~Hooft and M.~J.~G. Veltman, {\it {One loop divergencies in the theory of
  gravitation}},  {\em Ann. Inst. H. Poincare Phys. Theor. A} {\bf 20} (1974)
  69--94.

\bibitem{Goroff:1985th}
M.~H. Goroff and A.~Sagnotti, {\it {The Ultraviolet Behavior of Einstein
  Gravity}},  {\em Nucl. Phys. B} {\bf 266} (1986) 709--736.

\bibitem{Fermi:1933jpa}
E.~Fermi, {\it {Tentativo di una teoria dell'emissione dei raggi beta}},  {\em
  Ric. Sci.} {\bf 4} (1933) 491--495.

\bibitem{Anselmi:2005vk}
D.~Anselmi, {\it {Infinite reduction of couplings in non-renormalizable quantum
  field theory}},  {\em JHEP} {\bf 08} (2005) 029,
  [\href{http://arxiv.org/abs/hep-th/0503131}{{\tt hep-th/0503131}}].

\bibitem{Anselmi:2002ge}
D.~Anselmi, {\it {Absence of higher derivatives in the renormalization of
  propagators in quantum field theories with infinitely many couplings}},  {\em
  Class. Quant. Grav.} {\bf 20} (2003) 2355--2378,
  [\href{http://arxiv.org/abs/hep-th/0212013}{{\tt hep-th/0212013}}].

\bibitem{Anselmi:2013wha}
D.~Anselmi, {\it {Properties Of The Classical Action Of Quantum Gravity}},
  {\em JHEP} {\bf 05} (2013) 028, [\href{http://arxiv.org/abs/1302.7100}{{\tt
  arXiv:1302.7100}}].

\bibitem{Weinberg:1980gg}
S.~Weinberg, {\em {Ultraviolet divergences in quantum theories of
  gravitation}}, pp.~790--831.
\newblock 1980.

\bibitem{Lauscher:2001ya}
O.~Lauscher and M.~Reuter, {\it {Ultraviolet fixed point and generalized flow
  equation of quantum gravity}},  {\em Phys. Rev. D} {\bf 65} (2002) 025013,
  [\href{http://arxiv.org/abs/hep-th/0108040}{{\tt hep-th/0108040}}].

\bibitem{Stelle:1976gc}
K.~S. Stelle, {\it {Renormalization of Higher Derivative Quantum Gravity}},
  {\em Phys. Rev. D} {\bf 16} (1977) 953--969.

\bibitem{Julve:1978xn}
J.~Julve and M.~Tonin, {\it {Quantum Gravity with Higher Derivative Terms}},
  {\em Nuovo Cim. B} {\bf 46} (1978) 137--152.

\bibitem{Fradkin:1981iu}
E.~S. Fradkin and A.~A. Tseytlin, {\it {Renormalizable asymptotically free
  quantum theory of gravity}},  {\em Nucl. Phys. B} {\bf 201} (1982) 469--491.

\bibitem{Kawasaki:1981gk}
S.~Kawasaki and T.~Kimura, {\it {A Possible Mechanism of Ghost Confinement in a
  Renormalizable Quantum Gravity}},  {\em Prog. Theor. Phys.} {\bf 65} (1981)
  1767.

\bibitem{Tomboulis:1983sw}
E.~T. Tomboulis, {\it {Unitarity in Higher Derivative Quantum Gravity}},  {\em
  Phys. Rev. Lett.} {\bf 52} (1984) 1173.

\bibitem{Avramidi:1985ki}
I.~G. Avramidi and A.~O. Barvinsky, {\it {Asymptotic freedom in higher
  derivative quantum gravity}},  {\em Phys. Lett. B} {\bf 159} (1985) 269--274.

\bibitem{Starobinsky:1980te}
A.~A. Starobinsky, {\it {A New Type of Isotropic Cosmological Models Without
  Singularity}},  {\em Phys. Lett. B} {\bf 91} (1980) 99--102.

\bibitem{Lee:1969fy}
T.~D. Lee and G.~C. Wick, {\it {Negative Metric and the Unitarity of the S
  Matrix}},  {\em Nucl. Phys. B} {\bf 9} (1969) 209--243.

\bibitem{Lee:1970iw}
T.~D. Lee and G.~C. Wick, {\it {Finite Theory of Quantum Electrodynamics}},
  {\em Phys. Rev. D} {\bf 2} (1970) 1033--1048.

\bibitem{Modesto:2015ozb}
L.~Modesto and I.~L. Shapiro, {\it {Superrenormalizable quantum gravity with
  complex ghosts}},  {\em Phys. Lett. B} {\bf 755} (2016) 279--284,
  [\href{http://arxiv.org/abs/1512.07600}{{\tt arXiv:1512.07600}}].

\bibitem{Modesto:2016ofr}
L.~Modesto, {\it {Super-renormalizable or finite Lee\textendash{}Wick quantum
  gravity}},  {\em Nucl. Phys. B} {\bf 909} (2016) 584--606,
  [\href{http://arxiv.org/abs/1602.02421}{{\tt arXiv:1602.02421}}].

\bibitem{Nelson:1972vbp}
C.~A. Nelson and E.~C.~G. Sudarshan, {\it {Quantum field theories with shadow
  states. i. soluble models}},  {\em Phys. Rev. D} {\bf 6} (1972) 3658--3678.

\bibitem{Sudarshan:1972mwu}
E.~C.~G. Sudarshan and C.~A. Nelson, {\it {Quantum field theories with shadow
  states. ii. low-energy pion-nucleon scattering}},  {\em Phys. Rev. D} {\bf 6}
  (1972) 3678--3688.

\bibitem{Nelson:1972zq}
C.~A. Nelson, {\it {Physical unitarization of indefinite-metric theories by
  shadow state summation, bjorken scaling and light quarks}},  {\em Lett. Nuovo
  Cim.} {\bf 4S2} (1972) 913--918.

\bibitem{Kuzmin:1989sp}
Y.~V. Kuzmin, {\it {THE CONVERGENT NONLOCAL GRAVITATION. (IN RUSSIAN)}},  {\em
  Sov. J. Nucl. Phys.} {\bf 50} (1989) 1011--1014.

\bibitem{Modesto:2017sdr}
L.~Modesto and L.~Rachwa\l{}, {\it {Nonlocal quantum gravity: A review}},  {\em
  Int. J. Mod. Phys. D} {\bf 26} (2017), no.~11 1730020.

\bibitem{Donoghue:2019fcb}
J.~F. Donoghue and G.~Menezes, {\it {Unitarity, stability and loops of unstable
  ghosts}},  {\em Phys. Rev. D} {\bf 100} (2019), no.~10 105006,
  [\href{http://arxiv.org/abs/1908.02416}{{\tt arXiv:1908.02416}}].

\bibitem{Veltman:1963th}
M.~J.~G. Veltman, {\it {Unitarity and causality in a renormalizable field
  theory with unstable particles}},  {\em Physica} {\bf 29} (1963) 186--207.

\bibitem{Anselmi:2018ibi}
D.~Anselmi and M.~Piva, {\it {The Ultraviolet Behavior of Quantum Gravity}},
  {\em JHEP} {\bf 05} (2018) 027, [\href{http://arxiv.org/abs/1803.07777}{{\tt
  arXiv:1803.07777}}].

\bibitem{Kluberg-Stern:1974nmx}
H.~Kluberg-Stern and J.~B. Zuber, {\it {Renormalization of Nonabelian Gauge
  Theories in a Background Field Gauge. 1. Green Functions}},  {\em Phys. Rev.
  D} {\bf 12} (1975) 482--488.

\bibitem{Kluberg-Stern:1975ebk}
H.~Kluberg-Stern and J.~B. Zuber, {\it {Renormalization of Nonabelian Gauge
  Theories in a Background Field Gauge. 2. Gauge Invariant Operators}},  {\em
  Phys. Rev. D} {\bf 12} (1975) 3159--3180.

\bibitem{Anselmi:2015niw}
D.~Anselmi, {\it {Background field method and the cohomology of
  renormalization}},  {\em Phys. Rev. D} {\bf 93} (2016), no.~6 065034,
  [\href{http://arxiv.org/abs/1511.01244}{{\tt arXiv:1511.01244}}].

\bibitem{tHooft:1973bhk}
G.~'t~Hooft, {\it {An algorithm for the poles at dimension four in the
  dimensional regularization procedure}},  {\em Nucl. Phys. B} {\bf 62} (1973)
  444--460.

\bibitem{Anselmi:2016fid}
D.~Anselmi, {\it {Aspects of perturbative unitarity}},  {\em Phys. Rev. D} {\bf
  94} (2016) 025028, [\href{http://arxiv.org/abs/1606.06348}{{\tt
  arXiv:1606.06348}}].

\bibitem{Anselmi:2021hab}
D.~Anselmi, {\it {Diagrammar of physical and fake particles and spectral
  optical theorem}},  \href{http://arxiv.org/abs/2109.06889}{{\tt
  arXiv:2109.06889}}.

\bibitem{Anselmi:2020tqo}
D.~Anselmi, {\it {The quest for purely virtual quanta: fakeons versus
  Feynman-Wheeler particles}},  {\em JHEP} {\bf 03} (2020) 142,
  [\href{http://arxiv.org/abs/2001.01942}{{\tt arXiv:2001.01942}}].

\bibitem{Anselmi:2018kgz}
D.~Anselmi, {\it {Fakeons And Lee-Wick Models}},  {\em JHEP} {\bf 02} (2018)
  141, [\href{http://arxiv.org/abs/1801.00915}{{\tt arXiv:1801.00915}}].

\bibitem{Melis:2022tqz}
A.~Melis and M.~Piva, {\it {One-Loop Integrals for Purely Virtual Particles}},
  \href{http://arxiv.org/abs/2209.05547}{{\tt arXiv:2209.05547}}.

\bibitem{Anselmi:2021icc}
D.~Anselmi, K.~Kannike, C.~Marzo, L.~Marzola, A.~Melis, K.~M\"u\"ursepp,
  M.~Piva, and M.~Raidal, {\it {Phenomenology of a Fake Inert Doublet Model}},
  \href{http://arxiv.org/abs/2104.02071}{{\tt arXiv:2104.02071}}.

\bibitem{Anselmi:2018yct}
D.~Anselmi, {\it {On the nature of the Higgs boson}},  {\em Mod. Phys. Lett. A}
  {\bf 34} (2019), no.~16 1950123, [\href{http://arxiv.org/abs/1811.02600}{{\tt
  arXiv:1811.02600}}].

\bibitem{PivaMelis}
A.~Melis and M.~Piva {\em In preparation}.

\bibitem{Anselmi:2018tmf}
D.~Anselmi and M.~Piva, {\it {Quantum Gravity, Fakeons And Microcausality}},
  {\em JHEP} {\bf 11} (2018) 021, [\href{http://arxiv.org/abs/1806.03605}{{\tt
  arXiv:1806.03605}}].

\bibitem{Anselmi:2019rxg}
D.~Anselmi, {\it {Fakeons and the classicization of quantum gravity: the FLRW
  metric}},  {\em JHEP} {\bf 04} (2019) 061,
  [\href{http://arxiv.org/abs/1901.09273}{{\tt arXiv:1901.09273}}].

\bibitem{Anselmi:2020lpp}
D.~Anselmi, E.~Bianchi, and M.~Piva, {\it {Predictions of quantum gravity in
  inflationary cosmology: effects of the Weyl-squared term}},  {\em JHEP} {\bf
  07} (2020) 211, [\href{http://arxiv.org/abs/2005.10293}{{\tt
  arXiv:2005.10293}}].

\bibitem{Anselmi:2020neq}
D.~Anselmi, {\it {High-order corrections to inflationary perturbation spectra
  in quantum gravity}},  {\em JCAP} {\bf 02} (2021) 029,
  [\href{http://arxiv.org/abs/2010.04739}{{\tt arXiv:2010.04739}}].

\bibitem{Anselmi:2020shx}
D.~Anselmi, {\it {Cosmic inflation as a renormalization-group flow: the running
  of power spectra in quantum gravity}},  {\em JCAP} {\bf 01} (2021) 048,
  [\href{http://arxiv.org/abs/2007.15023}{{\tt arXiv:2007.15023}}].

\bibitem{Anselmi:2021rye}
D.~Anselmi, F.~Fruzza, and M.~Piva, {\it {Renormalization-group techniques for
  single-field inflation in primordial cosmology and quantum gravity}},  {\em
  Class. Quant. Grav.} {\bf 38} (2021), no.~22 225011,
  [\href{http://arxiv.org/abs/2103.01653}{{\tt arXiv:2103.01653}}].

\bibitem{Anselmi:2021dag}
D.~Anselmi, {\it {Perturbation spectra and renormalization-group techniques in
  double-field inflation and quantum gravity cosmology}},  {\em JCAP} {\bf 07}
  (2021) 037, [\href{http://arxiv.org/abs/2105.05864}{{\tt arXiv:2105.05864}}].

\bibitem{Planck:2018jri}
{\bf Planck} Collaboration, Y.~Akrami et~al., {\it {Planck 2018 results. X.
  Constraints on inflation}},  {\em Astron. Astrophys.} {\bf 641} (2020) A10,
  [\href{http://arxiv.org/abs/1807.06211}{{\tt arXiv:1807.06211}}].

\bibitem{BICEP:2021xfz}
{\bf BICEP, Keck} Collaboration, P.~A.~R. Ade et~al., {\it {Improved
  Constraints on Primordial Gravitational Waves using Planck, WMAP, and
  BICEP/Keck Observations through the 2018 Observing Season}},  {\em Phys. Rev.
  Lett.} {\bf 127} (2021), no.~15 151301,
  [\href{http://arxiv.org/abs/2110.00483}{{\tt arXiv:2110.00483}}].

\bibitem{Hazumi:2019lys}
M.~Hazumi et~al., {\it {LiteBIRD: A Satellite for the Studies of B-Mode
  Polarization and Inflation from Cosmic Background Radiation Detection}},
  {\em J. Low Temp. Phys.} {\bf 194} (2019), no.~5-6 443--452.

\end{thebibliography}\endgroup

\end{document}